\documentclass[twocolumn]{aastex631}

\newcommand{\oiii}{[\ion{O}{3}]}

\usepackage{comment}
\usepackage{amsmath}
\usepackage{CJKutf8}

\begin{document}

%\title{A SPectroscopic survey of biased halos In the Reionization Era (ASPIRE): JWST Reveals Reionization Completes Earlier around \oiii\ Emitters}

\title{A SPectroscopic survey of biased halos In the Reionization Era (ASPIRE): JWST Supports Earlier Reionization around \oiii\ Emitters}

\author[0000-0002-5768-738X]{Xiangyu Jin}
\affiliation{Steward Observatory, University of Arizona, 933 N. Cherry Ave., Tucson, AZ 85719, USA}

\author[0000-0001-5287-4242]{Jinyi Yang}
\affiliation{Steward Observatory, University of Arizona, 933 N. Cherry Ave., Tucson, AZ 85719, USA}

\author[0000-0003-3310-0131] {Xiaohui Fan}
\affiliation{Steward Observatory, University of Arizona, 933 N. Cherry Ave., Tucson, AZ 85719, USA}

\author[0000-0002-7633-431X]{Feige Wang}
\affiliation{Steward Observatory, University of Arizona, 933 N. Cherry Ave., Tucson, AZ 85719, USA}

\author[0000-0001-6874-1321]{Koki Kakiichi}
\affiliation{Cosmic Dawn Center (DAWN), Niels Bohr Institute, University of Copenhagen, Jagtvej 128, København N, DK-2200, Denmark}

\author[0000-0001-5492-4522]{Romain~A.~Meyer}
\affiliation{Department of Astronomy, University of Geneva, Chemin Pegasi 51, 1290 Versoix, Switzerland}

\author[0000-0003-2344-263X]{George D. Becker}
\affiliation{Department of Physics \& Astronomy, University of California, Riverside, CA 92521, USA}

\author[0000-0002-3983-6484]{Siwei Zou}
\affiliation{Chinese Academy of Sciences South America Center for Astronomy, National Astronomical Observatories, CAS, Beijing 100101, China}
\affiliation{Department of Astronomy, Tsinghua University, Beijing 100084, China} 

\author[0000-0002-2931-7824]{Eduardo Ba\~nados}
\affiliation{{Max Planck Institut f\"ur Astronomie, K\"onigstuhl 17, D-69117 Heidelberg, Germany}}

\author[0000-0002-6184-9097]{Jaclyn~B.~Champagne}
\affiliation{Steward Observatory, University of Arizona, 933 N. Cherry Ave., Tucson, AZ 85719, USA}

\author[0000-0003-3693-3091]{Valentina D’Odorico}
\affiliation{INAF - Osservatorio Astronomico, via G.B. Tiepolo, 11, I-34143 Trieste, Italy}
\affiliation{Scuola Normale Superiore, P.zza dei Cavalieri, I-56126 Pisa, Italy}
\affiliation{IFPU - Institute for Fundamental Physics of the Universe, via Beirut 2, I-34151 Trieste, Italy}

\author[0000-0002-5367-8021]{Minghao Yue}
\affiliation{MIT Kavli Institute for Astrophysics and Space Research, 77 Massachusetts Ave., Cambridge, MA 02139, USA}

\author[0000-0001-8582-7012]{Sarah E. I. Bosman}
\affiliation{Max Planck Institut f\"{u}r Astronomie, K\"{o}nigstuhl 17, D-69117 Heidelberg, Germany}
\affiliation{Institut f\"{u}r Theoretische Physik, Universit\"{a}t Heidelberg, Philosophenweg 16, D-69120 Heidelberg, Germany}

\author[0000-0001-8467-6478]{Zheng Cai}
\affiliation{Department of Astronomy, Tsinghua University, Beijing 100084, China}

\author[0000-0003-2895-6218]{Anna-Christina Eilers}
\affiliation{MIT Kavli Institute for Astrophysics and Space Research, 77 Massachusetts Ave., Cambridge, MA 02139, USA}

\author[0000-0002-7054-4332]{Joseph F. Hennawi}
\affiliation{Department of Physics, Broida Hall, University of California, Santa Barbara, CA 93106-9530, USA}
\affiliation{Leiden Observatory, Leiden University, P.O. Box 9513, NL-2300 RA Leiden, The Netherlands}

\author[0000-0003-1470-5901]{Hyunsung D. Jun}
\affiliation{Department of Physics, Northwestern College, 101 7th St SW, Orange City, IA 51041, USA}
\affiliation{SNU Astronomy Research Center, Seoul National University, 1 Gwanak-ro, Gwanak-gu, Seoul 08826, Republic of Korea}

\author[0000-0001-6251-649X]{Mingyu Li}
\affiliation{Department of Astronomy, Tsinghua University, Beijing 100084, China}

\author[0000-0001-5951-459X]{Zihao Li}
\affiliation{Department of Astronomy, Tsinghua University, Beijing 100084, China}

\author[0000-0003-3762-7344]{Weizhe Liu \begin{CJK}{UTF8}{gbsn}(刘伟哲)\end{CJK}}
\affiliation{Steward Observatory, University of Arizona, 933 N. Cherry Ave., Tucson, AZ 85719, USA}

\author[0000-0003-4924-5941]{Maria Pudoka}
\affiliation{Steward Observatory, University of Arizona, 933 N. Cherry Ave., Tucson, AZ 85719, USA}

\author[0000-0001-5818-6838]{Sindhu Satyavolu}
\affiliation{Tata Institute of Fundamental Research, Homi Bhabha Road, Mumbai 400005, India}

\author[0000-0002-4622-6617]{Fengwu Sun}
\affiliation{Center for Astrophysics $|$ Harvard \& Smithsonian, 60 Garden St., Cambridge MA 02138 USA}
\affiliation{Steward Observatory, University of Arizona, 933 N. Cherry Ave., Tucson, AZ 85719, USA}

\author[0000-0003-0747-1780]{Wei Leong Tee}
\affiliation{Steward Observatory, University of Arizona, 933 N. Cherry Ave., Tucson, AZ 85719, USA}

\author[0000-0003-0111-8249]{Yunjing Wu}
\affiliation{Department of Astronomy, Tsinghua University, Beijing 100084, China}

%% Note that the \and command from previous versions of AASTeX is now
%% depreciated in this version as it is no longer necessary. AASTeX 
%% automatically takes care of all commas and "and"s between authors names.

%% AASTeX 6.31 has the new \collaboration and \nocollaboration commands to
%% provide the collaboration status of a group of authors. These commands 
%% can be used either before or after the list of corresponding authors. The
%% argument for \collaboration is the collaboration identifier. Authors are
%% encouraged to surround collaboration identifiers with ()s. The 
%% \nocollaboration command takes no argument and exists to indicate that
%% the nearby authors are not part of surrounding collaborations.

%% Mark off the abstract in the ``abstract'' environment. 
\begin{abstract}
Understanding when and how reionization happened is crucial for studying the early structure formation and the properties of first galaxies in the Universe. At $z>5.5$, the observed IGM optical depth shows a significant scatter, indicating an inhomogeneous reionization process. However, the nature of the inhomogeneous reionization remains debated. ASPIRE is a JWST Cycle 1 program that has spectroscopically identified $>400$ \oiii\ emitters in 25 quasar fields at $z>6.5$. %These \oiii\ emitters are sensitive tracers of large-scale galaxy density. 
Combined with deep ground-based optical spectroscopy of ASPIRE quasars, ASPIRE program provides the current largest sample for IGM-galaxy connection studies during cosmic reionization. We present the first results of IGM effective optical depth measurements around \oiii\ emitters using 14 ASPIRE quasar fields. 
We find the IGM transmission is tightly related with reionization-era galaxies to the extent that significant excess of Ly$\alpha$ transmission exists around \oiii\ emitters. We measure the stacked IGM effective optical depth of IGM patches associated with \oiii\ emitters and find they reach the same IGM effective optical depth at least ${\rm d}z\sim0.1$ ahead of those IGM patches where no \oiii\ emitters are detected, supporting earlier reionization around \oiii\ emitters. 
Our results indicate an enhancement in IGM Ly$\alpha$ transmission around \oiii\ emitters at scales beyond 25$\;\!h^{-1}$~cMpc, consistent with the predicted topology of reionization from fluctuating UV background (UVB) models. 

\end{abstract}

%% Keywords should appear after the \end{abstract} command. 
%% The AAS Journals now uses Unified Astronomy Thesaurus concepts:
%% https://astrothesaurus.org
%% You will be asked to selected these concepts during the submission process
%% but this old "keyword" functionality is maintained in case authors want
%% to include these concepts in their preprints.

%% From the front matter, we move on to the body of the paper.
%% Sections are demarcated by \section and \subsection, respectively.
%% Observe the use of the LaTeX \label
%% command after the \subsection to give a symbolic KEY to the
%% subsection for cross-referencing in a \ref command.
%% You can use LaTeX's \ref and \label commands to keep track of
%% cross-references to sections, equations, tables, and figures.
%% That way, if you change the order of any elements, LaTeX will
%% automatically renumber them.
%%
%% We recommend that authors also use the natbib \citep
%% and \citet commands to identify citations.  The citations are
%% tied to the reference list via symbolic KEYs. The KEY corresponds
%% to the KEY in the \bibitem in the reference list below. 

\section{Introduction} \label{sec:intro}

Cosmic reionization was the last major phase transition of atomic hydrogen from neutral to ionized in the intergalactic medium (IGM). 
During cosmic reionization, ionized regions, created and powered by UV bright sources, gradually grew and overlapped in the IGM. Understanding when and how reionization happened not only reveals properties of the first luminous sources in the universe, but also provides crucial information of the early structure formation and large scale structure in the early universe \citep{Finkelstein2019ApJ,Robertson2022ARAA}. 

Observations have revealed that reionization ended at $z<6$ through various astrophysical sources. The Thomson scattering optical depth observed from the CMB observations imply a midpoint of reionization around $z\sim7.7$ \citep{Planck2020AA}. At $z\gtrsim6$, IGM Lyman $\alpha$ (Ly$\alpha$) damping wing profiles -- strong absorption caused by significantly neutral gas -- are widely observed among luminous quasars \cite[e.\,g.,\,][]{Greig2017MNRAS,Greig2019MNRAS,Greig2022MNRAS,Greig2024arXiv,Banados2018Natur,Davies2018ApJb,Wang2020ApJ,Yang2020ApJa,Durovcikova2024arXiv}, in galaxies (\citealt{Bunker2023AA,Umeda2023arXiv}, but see \citealt{Heintz2023arXiv,Heintz2024arXiv,Keating2023arXiv}), and in gamma-ray burst afterglows (\citealt{Totani2016PASJ}, but see \citealt{Hartoog2015A&A}). 
%(note the redshift is lower and the neutral fraction is also lower). 
The discovery of Gunn-Peterson troughs, where the observed flux in the Ly$\alpha$ forest is consistent with zero, is smoking gun evidence for an ongoing reionization at $z\sim6$ \citep{GP1965ApJ,Becker2001AJ}. At $z<6$, prominent transmission spikes, as opposed to the dark absorption troughs, are emergent in the Ly$\alpha$ forest \citep{McGreer2011MNRAS,McGreer2015MNRAS,Zhu2021ApJ,Zhu2022ApJ,Jin2023ApJ}. %From $z>6$ to $z<6$, the Ly$\alpha$ visibility of galaxies increases {\color{blue}{+reference}}. 

Meanwhile, the Ly$\alpha$ visibility of galaxies increases rapidly from $z>6$ to $z<6$ \cite[e.\,g.,\,][]{Stark2010MNRAS,Stark2011ApJ,Caruana2014MNRAS,Pentericci2014ApJ,Schenker2014ApJ,Debarros2017AA,Mason2018ApJ,Mason2019MNRAS}. 
%as well as the fast evolution of the observed clustering of Ly$\alpha$ emitters \citep{SM2015MNRAS,Ouchi2018PASJ}, and Ly$\alpha$ luminosity function \citep{Morales2021ApJ} {\color{blue}{+reference}}. 
Those observations support a rapid evolution in the IGM neutral fraction at $6<z<8$. Accumulating observational evidence shows that reionization might still be ongoing at $z<6$ \cite[for a review, see][]{Fan2023ARAA}, % in the format of fluctuating UVB or neutral islands
including recent ionizing photon mean free path measurements showing a dramatic evolution at $5<z<6$ \citep{Becker2021MNRAS,Zhu2023ApJ}, the existence of dark regions in both quasar Ly$\alpha$ and Ly$\beta$ forests at $z<6$ \citep{Zhu2021ApJ,Zhu2022ApJ,Jin2023ApJ}, detections of damping wing absorption in the Ly$\alpha$ forest at $z<6$ \citep{Becker2024MNRAS,Zhu2024MNRAS,Spina2024AA}, and the large scatter in the observed IGM effective optical depth ($\tau_{\rm eff}$) distribution at $z\gtrsim5.4$ which cannot be explained by homogeneous ionizing background models \citep{Yang2020ApJb,Bosman2022MNRAS}. 
% {\color{blue}{+evidence of a late reionization: dark gaps, mean free path, Sarah's IGM work}}. 
%the Ly$\alpha$ emitting fraction, the Ly$\alpha$ equivalent width distribution, Ly$\alpha$ emitter clustering, Ly$\alpha$ luminosity function, are rapidly increasing. 

At $z\sim5.5$, although $\tau_{\rm eff}$ measured from Ly$\alpha$ forests suggests the IGM is highly ionized with an average neutral fraction $\lesssim10^{-4}$, 
% neutral fraction measurements: \citep{Fan2006AJ,Becker2015MNRAS,Yang2020ApJb, Bosman2022MNRAS}
the observed $\tau_{\rm eff}$ distribution (usually measured in a bin size of 50$\;\!h^{-1}$~cMpc), displays an increased scatter at $z>5.5$ \citep{Fan2006AJ,Becker2015MNRAS,Bosman2018MNRAS,Bosman2022MNRAS,Eilers2018ApJ,Yang2020ApJb}. The scatter in $\tau_{\rm eff}$ at $z>5.5$ reflects large-scale variations close to the end of reionization and implies a patchy and inhomogeneous reionization process. 
%To explain scatter in the IGM effective optical depth at $z>5.5$, several reionization models have proposed different topology of reionization. 
\citet{Fan2006AJ} propose that the large scatter is driven by the fluctuations in the UV ionizing background (UVB) while \citet{Lidz2006ApJ} argue that the scatter could be explained by density field fluctuations with a uniform ionizing background. \citet{Daloisio2015ApJ} suggest that variations in the IGM temperature can also explain the observed fluctuation in $\tau_{\rm eff}$. These reionization models predict distinct topology of reionization to the extent of the relation between the density field and large-scale IGM transmission. In the fluctuating UVB models, overdense regions of galaxies correspond to enhancement in the IGM transmission because galaxies enhance ionizing radiation field around them \citep{Davies2016MNRAS,Davies2018ApJa,Davies2024ApJ}. On the other hand, if large scale fluctuations are driven by density field or IGM temperature, overdense regions in the IGM correspond to the suppression of IGM transmission \cite[e.\,g.,\,][]{Davies2018ApJa}. 

%In addition, a few simulation studies have investigated the relation between narrow IGM Ly$\alpha$ transmission spikes and galaxies. \citet{Garaldi2019ApJ} find transmission spikes are mainly produced by highly ionized underdense regions in vicinity of UV sources in the IGM. However, \citet{Zhu2024arXiv} argue that transmission spikes are primarily generated by low gas density, while the contribution from galaxies to local ionizing background is minimal in producing transmission spikes. 

% density field fluctuation: overdense regions-> overdensity of neutral gas; IGM temperature: overdense regions get ionized earlier and thus have longer time to cool in the IGM, neutral hydrogen recombines around overdense regions.  

%{{\color{red}{\Large HERE}}} 
Previous observational studies have investigated the relation between galaxies and IGM transmission using ground-based observations. \citet{Becker2018ApJ,Christenson2021ApJ,Christenson2023ApJ,Ishimoto2022MNRAS} have utilized Hyper-Suprime Cam (HSC) on the Subaru Telescope to select Ly$\alpha$ emitters (LAEs) at $z\sim5.7$ and have investigated the relation between the surface number density of LAEs and $\tau_{\rm eff}$ using a few quasar sightlines. Significant underdensity of LAEs has been found around a few highly opaque IGM patches at $z\sim5.7$ \citep{Becker2018ApJ,Christenson2021ApJ}, consistent with the fluctuating UVB models. Using broad band selected Lyman-break galaxies (LBGs), \citet{Kashino2020ApJ} find an underdensity of LBGs at $5.5<z<5.9$ along the long dark trough at $z\sim5.7$ exhibited in the a quasar J0148+0600, also consistent with fluctuating UVB models. However, for highly transparent quasar sightlines, both overdensities \citep{Ishimoto2022MNRAS} and underdensities \citep{Christenson2023ApJ} of LAEs have been found, which may be in tension with either fluctuating UVB or IGM temperature models. As addressed in \citet{Christenson2023ApJ}, at lower redshift, LAEs are found to avoid high-density peaks in the IGM \cite[e.\,g.,\,][]{Kashikawa2007ApJ,Momose2021ApJ,Huang2022ApJ}, because Ly$\alpha$ emission can be suppressed in high density of neutral gas (\citealt{Huang2022ApJ}, see also \citealt{Tang2024arXiv}), or star formation activity is quenched by the strong ionizing background in the high-density regions \citep{Kashikawa2007ApJ,Lambert2024arXiv}. Therefore, other tracers of the density field are needed.  %{\color{blue}{LAEs might avoid high-density peaks in the IGM and the Lyman-break galaxies can have pretty uncertain redshift.+discussion in christenson 2023; Selection of LAEs and LBGs is limited to certain redshifts based on the filters}} 
In addition, \citet{Kakiichi2018MNRAS,Meyer2019MNRAS,Meyer2020MNRAS} measure the cross-correlation function between the quasar Ly$\alpha$ forest spectrum and various galaxy populations, including LBGs, faint galaxies traced by \ion{C}{4} absorbers, and LAEs identified in the quasar field, and find excess IGM transmission with $\sim3\sigma$ at $\gtrsim10$\,cMpc, suggesting enhanced IGM transmission is associated with galaxies. However, selecting high-redshift galaxies with ground-based observations is expensive and the number of galaxies identified in the quasar fields is also limited, resulting in difficulty of detecting significant signals in the IGM-galaxy cross-correlation function. It is thus preferred to use space-based observations to select enough high-redshift galaxies in quasar fields and investigate the relation between IGM transmission and galaxies in a statistical manner. 

\textit{James Webb Space Telescope (JWST)} has revolutionized identification of high-redshift galaxies by detecting their rest-optical emission lines. Utilizing JWST NIRCam Wide Field Slitless Spectroscopy (WFSS) mode, the Emission-line galaxies and Intergalactic Gas in the Epoch of Reionization (EIGER) program (PID-1243) targets six quasar fields at $6.0<z<7.1$.   \citet{Kashino2023ApJ} identify 117 \oiii\ emitters in one quasar field at $z\sim6.3$ from the EIGER program, and measure the IGM-galaxy cross correlation function based on 59 \oiii\ emitters located within the Ly$\alpha$ and Ly$\beta$ forest. 
%At $5.70<z<6.14$, a clear excess of IGM transmission is found at a distance of $\sim5$~cMpc from galaxies in one quasar field. 
A SPectroscopic survey of biased halos In the Reionization Era (ASPIRE, PID-2078) is a program that targets 25 quasar fields at $6.5<z<6.8$. Along with ground-based spectroscopy covering the quasar Ly$\alpha$ forest, ASPIRE provides the current largest sample of quasar fields for studying the connection between galaxies and IGM during the reionization era. In this paper, we investigate the relation between galaxies and IGM transmission in a statistical manner, utilizing 14 ASPIRE quasar fields. The paper is organized as follows: In Section \ref{sec:data}, we introduce the data reduction of JWST WFSS data, the selection of \oiii\ emitters, and the data reduction of ground-based optical spectroscopy. In Section \ref{sec:method}, we introduce the method to quantify the IGM transmission around \oiii\ emitters. We show the stacked IGM transmission around \oiii\ emitters in Section \ref{sec:result}. In Section \ref{sec:discussion}, we investigate the $\tau_{\rm eff}$ distribution around \oiii\ emitters at different redshifts and on various scales. We summarize the paper in Section \ref{sec:conclusion}. Throughout this paper, we adopt a flat $\Lambda$CDM cosmology with cosmological parameters $H_{0}=70~{\rm km\;\!s^{-1}\;\!Mpc^{-1}}$ and $\Omega_{m}=0.3$.

%Quasar absorption spectroscopy plays an important role in probing the process of cosmic reionization. {\color{blue}{add one sentence here}} {\color{blue}{Other quasar probes?}} Thanks to the progress made in the quasar survey \citep{Fan2023ARAA}, more than 500 quasars have been discovered at $z>5.3$, enabling the quasar absorption studies during cosmic reionization in a statistical manner. With a large sample of quasars, {\color{blue}{+reference}} investigate the IGM effective optical depth of quasar Ly$\alpha$ forests and find a rapid increase in the average IGM effective optical depth with redshift. Meanwhile, a large scatter in the IGM effective optical depth also appears between different quasar sightlines at $z>5.5$, indicating an inhomogenous reionziation process. 

\begin{comment}
\begin{itemize}
    \item discovery of Gunn-Peterson troughs
    \item Effective optical depth work
    \item damping wing; dark pixel; 
\end{itemize}

Timeline of reionization 

topology of reionization, why is it important to study it? 
\end{comment}

\section{Data Preparation} \label{sec:data}

To probe the connection between galaxies and IGM, we use data from ASPIRE program (PI: F. Wang) to select galaxies in the quasar fields through \oiii\ doublet emission lines. We present ground-based optical spectroscopy covering the Ly$\alpha$ forest of ASPIRE quasars to measure the IGM Ly$\alpha$ transmission. In this section, we summarize the data reduction details of JWST NIRCam observations and ground-based optical spectroscopy. 

\subsection{JWST NIRCam/WFSS Observations from ASPIRE Program}

ASPIRE is a program that targets 25 quasar fields at $6.5<z<6.8$ with NIRCam Imaging and Wide-field Slitless Spectroscopy \cite[WFSS,][]{Rieke2005SPIE,Greene2017JATIS}. The WFSS mode enables an efficient identification of reionization-era galaxies through their rest-optical emission lines \cite[e.g.,\,][]{Sun2022ApJ}. The overview of ASPIRE program can be found in \citet{Wang2023ApJ,Yang2023ApJ} and we briefly summarize it below. In each quasar field, the F356W WFSS observations in the long-wavelength (LW) channel were taken simultaneously with the F200W imaging in the short-wavelength (SW) channel with a total on-source time of 2834.5~s. Imaging (includes direct imaging and out-of-field imaging) in F115W and F356W were taken in each quasar field with a total exposure time from 1417.3s near the quasar, to 472.4s at the edge of the field. There is a single pointing in each quasar field and the target quasar is designed to be located at the Module A with an X offset of $-60.5''$ and a Y offset of $7.5''$. We refer readers to \citet{Wang2023ApJ} and \citet{Yang2023ApJ} for more details about the data reduction. 

Using ASPIRE program, H$\beta$ and \oiii\ emission lines of $z\sim5.4-7.0$ galaxies can be identified by F356W WFSS data \cite[e.\,g.,\,][Champagne et al., 2024a, b, submitted]{Wang2023ApJ,Yang2023ApJ,Wu2023ApJ,Zou2024ApJ}. We follow the procedure in \citet{Wang2023ApJ} to search for \oiii\ emitters. We first extract the 1D spectrum from WFSS data and place a median filter of 51 pixels
%510~\AA\ 
on the spectrum. We perform a peak finding on the extracted spectrum to search the emission lines with $>5\sigma$ detection. To search \oiii\ emitters, we then assume the detected emission line is \oiii$\lambda$5008 and search the spectra with \oiii$\lambda$4960 with $>2\sigma$ detection. Those selected line emitters are viewed as \oiii\ emitter candidates. We then perform a visual inspection on all \oiii\ emitter candidates and remove those that are likely selected due to source confusion and contaminant line emitters at low-redshift. More details regarding the detection of line-emitters and identification of \oiii\ emitters can be found in Wang et al., 2024 in preparation. %478 \oiii\ emitters have been identified in 23 quasar fields with a limiting luminosity down to $\sim 10^{42}~{\rm erg\;\!s^{-1}}$ {\color{red} XJin: might need update from Feige}. 

\subsection{Ground-based Optical Spectroscopy}
To measure the transmission in quasar Ly$\alpha$ forests, we utilize existing ground-based optical spectroscopy \citep{Yang2020ApJb,Dodorico2023MNRAS} of ASPIRE quasars, and collect new optical spectroscopy of ASPIRE quasars. Overall, we are able to analyze 14 ASPIRE quasar sightlines given by the current available data. Broad absorption line quasars are not included in the analysis. 
%6 highest S/N sightlines out of 16 quasars are being analyzed and presented in Kakiichi et al., 2024 in preparation. 
%{\color{red}{delete it? The rest of the quasar fields will be presented as the full ASPIRE sample later?}}. 
We present details of these optical spectroscopy in the following subsections, and summarize the optical observations and number of \oiii\ emitters identified in quasar fields in Table \ref{tab:o3info}.

\subsubsection{Archival Optical Spectroscopy}
J0224$-$4711
%J0923$+$0402, 
and J1526$-$2050
%J2132$+$1217 
are included in the ESO Large Program The Ultimate XSHOOTER Legacy Survey of Quasars at the Reionization Epoch (XQR-30, PI: V. D’Odorico) and J0226$+$0302 is included as part of enlarged XQR-30 sample \citep[E-XQR-30, ][]{Dodorico2023MNRAS}. XSHOOTER is an Echelle spectrograph on the Very Large Telescope (VLT) with a wavelength coverage of $300-2500$~nm covered by three arms UVB, VIS, and NIR. In the optical band covered by the VIS arm, XSHOOTER achieves a spectral resolution $R$ of $\sim8800$ with a $0.9''$ slit \citep{Vernet2011AA}. We download the E-XQR-30 reduced optical spectra from the Github repository\footnote{\url{https://github.com/XQR-30/Spectra}}. The reduced E-XQR-30 spectra are rebinned to 10~${\rm km\;\!s^{-1}}$. 

In addition, we collect reduced archival optical spectroscopy of J0109$-$3047, J0305$-$3150, J0910$+$1656, J1048$-$0109, J1104$+$2134, J2002$-$3013, J2102$-$1458, and J2232$+$2930 from \citet{Yang2020ApJb}, and J1129$+$1846 from \citet{Banados2021ApJ}.
% delete J0218 here; Use it GMOS spectrum instead
J0109$-$3047, J0305$-$3150, J1048$-$0109, J1129$+$1846, and J2232$+$2930 were observed with VLT/XSHOOTER. 
%J0218$+$0007, 
%J0910$+$1656 and 
J1104$+$2134 was observed with Keck/LRIS \citep{Oke1995PASP,Rockosi2010SPIE}, with a 600/10000 grating and an 1$''$ slit, resulting in a spectral resolution of $R\sim1900$. J2002$-$3013 was observed with the Gemini/GMOS-S with the R400 grating and a $0.5''$ slit \citep{Gimeno2016SPIE,Hook2004PASP}, with a spectral resolution of $R\sim1900$. J2102$-$1458 was observed with Keck/DEIMOS with the 830G grating and an $1''$ slit \citep{Faber2003SPIE}, resulting a spectral resolution of $R\sim3300$. We refer readers to \citet{Yang2020ApJb} for more data reduction details. 

\subsubsection{New Optical Spectroscopy}

Apart from archival data, we also include new optical spectroscopy from 6--10m telescopes. J0218$+$0007 and J0525$-$2406 were observed in the Gemini/GMOS-S queues with R400 with a central wavelength of 855~nm and 860~nm in 2019 with the R400 grating and a spectral resolution of $R\sim1900$ (PI: J. Yang). J0244$-$5008 was observed with Magellan Clay/LDSS3 with VPH-Red (grism) with an 1$''$ slit (PI: M. Yue). The central wavelength is 8000~${\rm \AA}$ and the spectral resolution is $R\sim1800$. %J0706$+$2921 was observed with Keck/DEIMOS with an OG550 filter and a 830G grating and a central wavelengths of 8500~${\rm \AA}$ and 8520~${\rm \AA}$ in Dec, 2019 (PI: X. Fan). 

% on Dec 01, 2019
%J1129$+$1846 was observed by VLT/XSHOOTER in 2019 \cite[PI: C. Mazzucchelli, ][]{Banados2021ApJ}. 

For the new optical spectroscopy, we use \texttt{PypeIt} v.1.14.0 \citep{Prochaska2020JOSS,Prochaska2020zndo} to perform data reduction for bias subtracting, flat-fielding, and flux calibration, following the standard procedure\footnote{\url{https://pypeit.readthedocs.io/en/release/cookbook.html}}. 

\begin{figure}[!h]
    \centering
    \includegraphics[width=0.45\textwidth]{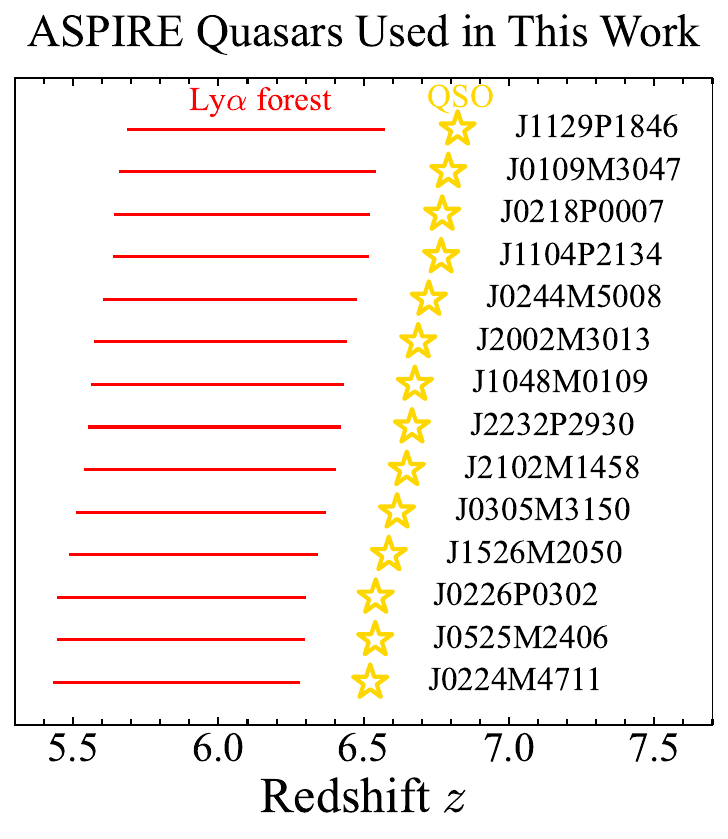}
    \caption{The redshift distribution (yellow) of ASPIRE quasars used in this work, in the ascending order of redshift, and the redshift range of the Ly$\alpha$ forest used in the analysis is denoted by the red horizontal line. We exclude the Ly$\alpha$ forest spectrum at $>1176~$\AA\ rest frame to exclude the possible emission from the quasar proximity zone and exclude the rest $\lambda<1040~{\rm \AA}$ Ly$\alpha$ forest to exclude the contamination from \ion{O}{6}$\lambda$~1033 emission and the Ly$\beta$ forest. The redshift range of ASPIRE quasars used in this work is from $z\sim6.5$ to $z\sim6.8$ and the corresponding Ly$\alpha$ absorption redshift range is from $z\sim5.4$ up to $z\sim6.6$.}
    \label{fig:all_quasar_sightlines}
\end{figure}

For each quasar sightline, following \citet{Yang2020ApJb}, we exclude a rest-frame wavelength range $<1040~{\rm \AA}$ in the Ly$\alpha$ forest to remove the contamination from Ly$\beta$ and \ion{O}{6} emission lines. We also exclude the Ly$\alpha$ forest spectrum at rest wavelength $>1176~{\rm \AA}$ to avoid possible emission from quasar proximity zone. The redshift distribution of ASPIRE quasars used in this work and the redshift coverage of their Ly$\alpha$ forest are shown in Figure \ref{fig:all_quasar_sightlines}. Following \citet{Yang2020ApJb} and \citet{Jin2023ApJ}, we use the least-squares method to fit the spectrum within rest-frame 1245-1285~\AA\ and 1310-1380~\AA\, by assuming a broken power-law with a break at 1000~\AA\ \citep{Shull2012ApJ}. The quasar intrinsic continuum flux is calculated using the best-fit power-law continuum. Transmitted flux in the Ly$\alpha$ forest ($T_{\rm Ly\alpha}$) is then calculated as the normalized flux $T_{\rm Ly\alpha}=F_{\rm obs}/F_{\rm cont}$, where $F_{\rm obs}$ is the observed flux in the Ly$\alpha$ forest and $F_{\rm cont}$ is the flux of the best-fit power-law continuum at the same wavelength. To remove the contamination from strong sky emission lines to IGM transmission, following 
\citet{Jin2023ApJ}, we mask pixels likely influenced by strong sky emission lines. 
%{\color{red} Those newly-reduced quasar spectra are presented in Figure xxx, together with \oiii\ emitters identified in the quasar fields.} 
%Figure \ref{fig:stacked_lya_forest_spectrum} shows the stacked Ly$\alpha$ forest spectrum of 14 quasar sightlines.

As the depth of the optical spectroscopy varies by quasar sightlines, we measure the $2\sigma$ limiting optical depth $\tau_{\rm lim, 2\sigma}$ of each quasar sightline to quantify the depth of the optical spectroscopy. A higher $\tau_{\rm lim, 2\sigma}$ indicates the quasar spectrum can detect weaker IGM transmission. We adopt the formula $\tau_{\rm lim, 2\sigma}=-{\rm ln}\langle2\sigma/F_{\rm cont}\rangle$ to calculate $\tau_{\rm lim, 2\sigma}$, where $\sigma$ is the spectral uncertainty and $F_{\rm cont}$ is the continuum flux calculated from the best-fit power-law continuum described above. The $\tau_{\rm lim, 2\sigma}$ is measured within a 50$\;\!h^{-1}$~cMpc bin centered at $z=6.0$, and 
the $\tau_{\rm lim, 2\sigma}$ at $z\sim6$ of 14 quasar sightlines used in this work is $\sim3.7-6.7$. Out of 14 ASPIRE quasars, one-third of the quasar sightlines with highest S/N and the corresponding IGM-galaxy cross-correlation function will be presented in Kakiichi et al., 2024 in preparation.

\begin{deluxetable*}{ccccccc}\centering
\tabletypesize{\scriptsize}
\tablecaption{Summary of ASPIRE Quasars Used in This Work}
\tablewidth{2\columnwidth}
\tablehead{\colhead{Quasar Name} &
\colhead{Quasar Redshift} & \colhead{\# of O3Es} & \colhead{Telescope/Instrument} & \colhead{Exposure time (hrs)} & \colhead{Reference} & \colhead{$\tau_{\rm lim, 2\sigma}\;\!(z=6.0)$}}
\colnumbers
\startdata
%J0100p2802 &  & 117 \\
J0109$-$3047 & 6.7909 & 23 & VLT/X-Shooter & 6.0 & \citet{Yang2020ApJb} & 4.60\\
J0218$+$0007 & 6.77 & 31 & Gemini/GMOS & 4.0 & New data & 4.72\\
J0224$-$4711 & 6.5222 & 17 & VLT/X-Shooter & 8.6 & \citet{Dodorico2023MNRAS} & 6.32\\
J0226$+$0302 & 6.5412 & 34 & VLT/X-Shooter & 6.5 & \citet{Dodorico2023MNRAS} & 6.65 \\
%J0229$-$0808 & 6.7249 & 20 & & &  & \\
J0244$-$5008 & 6.724 & 32 & Magellan/LDSS3 & 2.3 & New data & 5.79\\
J0305$-$3150 & 6.6145 & 50 & VLT/X-Shooter & 4.0 &  \citet{Yang2020ApJb}  & 5.02 \\
%J0430$-$1445 & 6.7142 & 18 & & & &  & \\
J0525$-$2406 & 6.5397 & 14 & Gemini/GMOS & 2.7 & New data & 3.70 \\
%%%% BAL % J0706$+$2921 & 6.6037 & 15 & Keck/DEIMOS & 4.9 & New data & 5.38 \\
%J0910$+$1656 & 6.7289 &  -  & LRIS & 2 & \citet{Yang2020ApJb} &  &  \\
%J0910$-$0414 & 6.6363 & 33 & & & &  & \\
%J0921$+$0007 & 6.5646 & 33 & & & &  & \\
%%%% BAL % J0923$+$0402 & 6.633 & 25 & VLT/X-Shooter & 12.0 & \citet{Dodorico2023MNRAS}  & 6.49 \\
%J0923$+$0753 & 6.6817 & 7 & & & &  & \\
J1048$-$0109 & 6.6759 & 18 & VLT/X-Shooter & 1.3 &  \citet{Yang2020ApJb} & 4.48 \\
%J1058$+$2930 & 6.5846 & 31 & & & & & \\
J1104$+$2134 & 6.7662 & 8 & Keck/LRIS & 2.0 & \citet{Yang2020ApJb} & 5.78 \\
%J1110$-$1329 & 6.5148 & 9 & & & &  & \\
J1129$+$1846 & 6.823 & 5 & VLT/X-Shooter & 4.5 & \citet{Banados2021ApJ}  & 5.48 \\
J1526$-$2050 & 6.5864 & 20 & VLT/X-Shooter & 12.2 & \citet{Dodorico2023MNRAS} & 6.52 \\
J2002$-$3013 & 6.6876 & 12 & Gemini/GMOS & 2.3 & \citet{Yang2020ApJb} & 5.49\\
J2102$-$1458 & 6.648 & 10 & Keck/DEIMOS & 3.3 & \citet{Yang2020ApJb} & 4.91  \\
%J2132$+$1217 & 6.585 & - & X-Shooter &  & \citet{Dodorico2023MNRAS}  &  & \\ 
J2232$+$2930 & 6.666 & 13 & VLT/X-Shooter & 4.0 & \citet{Yang2020ApJb}  & 5.53\\
\enddata
\tablecomments{(1) Quasar name, in the ascending order of R.A; (2) Quasar spectroscopic redshift; (3) Number of $5.4<z<7.0$ \oiii\ emitters identified in the quasar fields; (4) Telescope and instrument of the optical spectroscopy; (5) Exposure time in hours (hrs); (6) Reference of the optical spectroscopy reduction; (7) $2\sigma$ limiting optical depth, calculated in a $50\;\!h^{-1}~$cMpc centered bin at $z\sim6.0$. }
\end{deluxetable*}\label{tab:o3info}

\section{Methods} \label{sec:method}
%ensuring that transmissive IGM patches will be easily identified among the datasets. 

\begin{figure*}[!ht]
    \centering
    \includegraphics[width=0.95\textwidth]{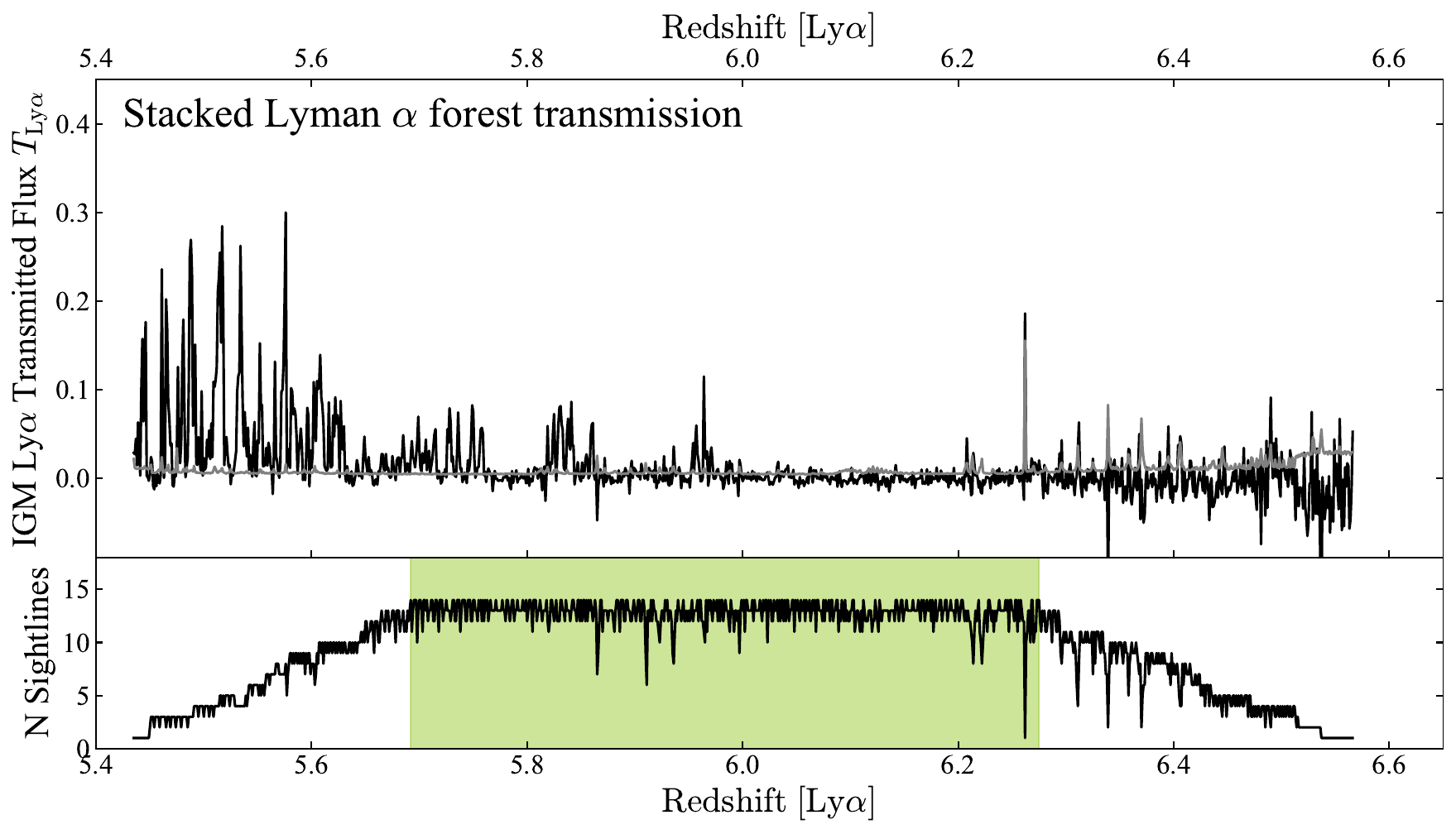}
    \caption{\textit{Top} -- The stacked Lyman $\alpha$ forest spectrum (black) of 14 ASPIRE quasar sightlines used in our analysis, adopting a redshift grid of ${\rm d}z=0.001$ from $z=5.435$ to $z=6.568$. The spectral uncertainty of the stacked spectrum is shown in grey. \textit{Bottom} -- The number of quasar line of sights (LOSs) available in each pixel of the redshift grid. Nearly all quasars used in this work covers the $z\sim5.70-6.27$ Ly$\alpha$ forest (the green shaded region). Some pixels are strongly influenced by sky OH emission lines, therefore, they have been removed in the analysis, resulting in troughs shown in the number of quasar LOSs.}
    \label{fig:stacked_lya_forest_spectrum}
\end{figure*}

\begin{figure}[!ht]
    \centering
    \includegraphics[width=0.42\textwidth]{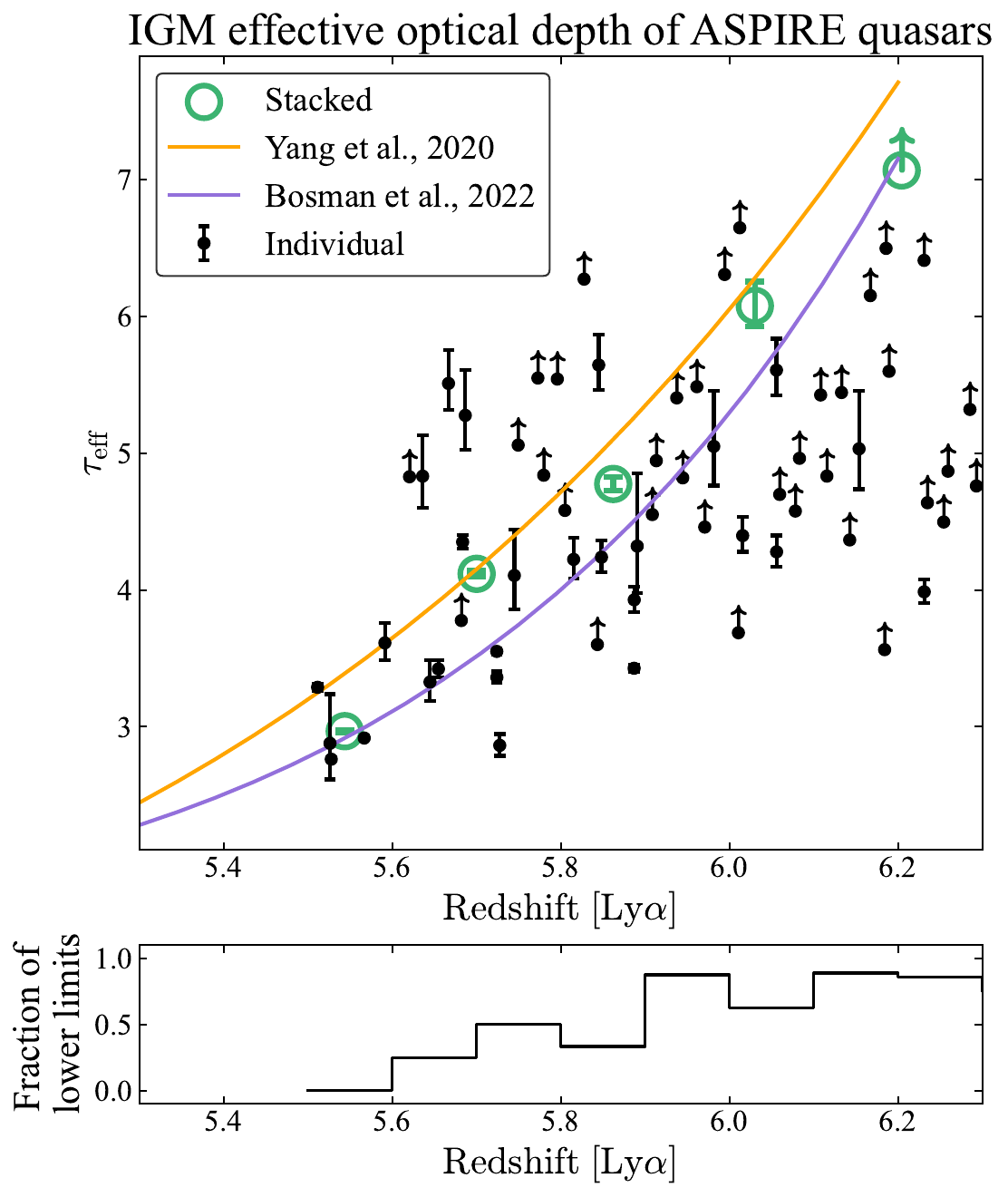}
    \caption{\textit{Top} -- The IGM Ly$\alpha$ effective optical depth $\tau_{\rm eff}$ of 14 ASPIRE quasar sightlines used in this work and the $\tau_{\rm eff}$ of the stacked Ly$\alpha$ forest spectrum in a bin size of 50~$h^{-1}\;\!{\rm cMpc}$. All the lower limits in the $\tau_{\rm eff}$ are shown in upwards arrows. As a comparison, we show the best-fit $\tau_{\rm eff}$ redshift evolution from \citet[][orange]{Yang2020ApJb} and \citet[][purple]{Bosman2022MNRAS}, derived from larger samples of 32 quasars and 67 quasars. \textit{Bottom} -- The fraction of $2\sigma$ lower limits in the $\tau_{\rm eff}$ measurements within a redshift bin of ${\rm d}z=0.1$. Beyond $z\sim6.1$, more than 75\% measurements are $2\sigma$ lower limits.}
    \label{fig:stack_tau_eff}
\end{figure}

\begin{comment}
\begin{figure*}[!ht]
    \centering
    \includegraphics[width=0.9\textwidth]{figures/illustrative_fig.png}
    \caption{An illustrative figure for measuring the IGM effective optical depth around the \oiii\ emitters. For each \oiii\ emitter detected in the quasar field, we put a sphere of a certain radius (so-called ``influence radius") centered at the \oiii\ emitter. If the sphere intersects the quasar sightline (i.\,e.,\, $r_{\perp}<$ influence radius), we then measure the IGM Ly$\alpha$ effective optical depth over the path length ($\sim2\times\sqrt{({\rm impact\;\!parameter})^2-r_{\perp}^2}$) enclosed by the sphere. If the sphere centered at an \oiii\ emitter does not intersect the quasar sightline, then we think the \oiii\ emitter is too far to impact the IGM traced by the quasar Ly$\alpha$ forest and thus to exclude it from our analysis {\color{red} Re-word}.}
    \label{fig:illustrative}
\end{figure*}
\end{comment}

\begin{figure*}[!ht]
    \centering
    \includegraphics[width=0.9\textwidth]{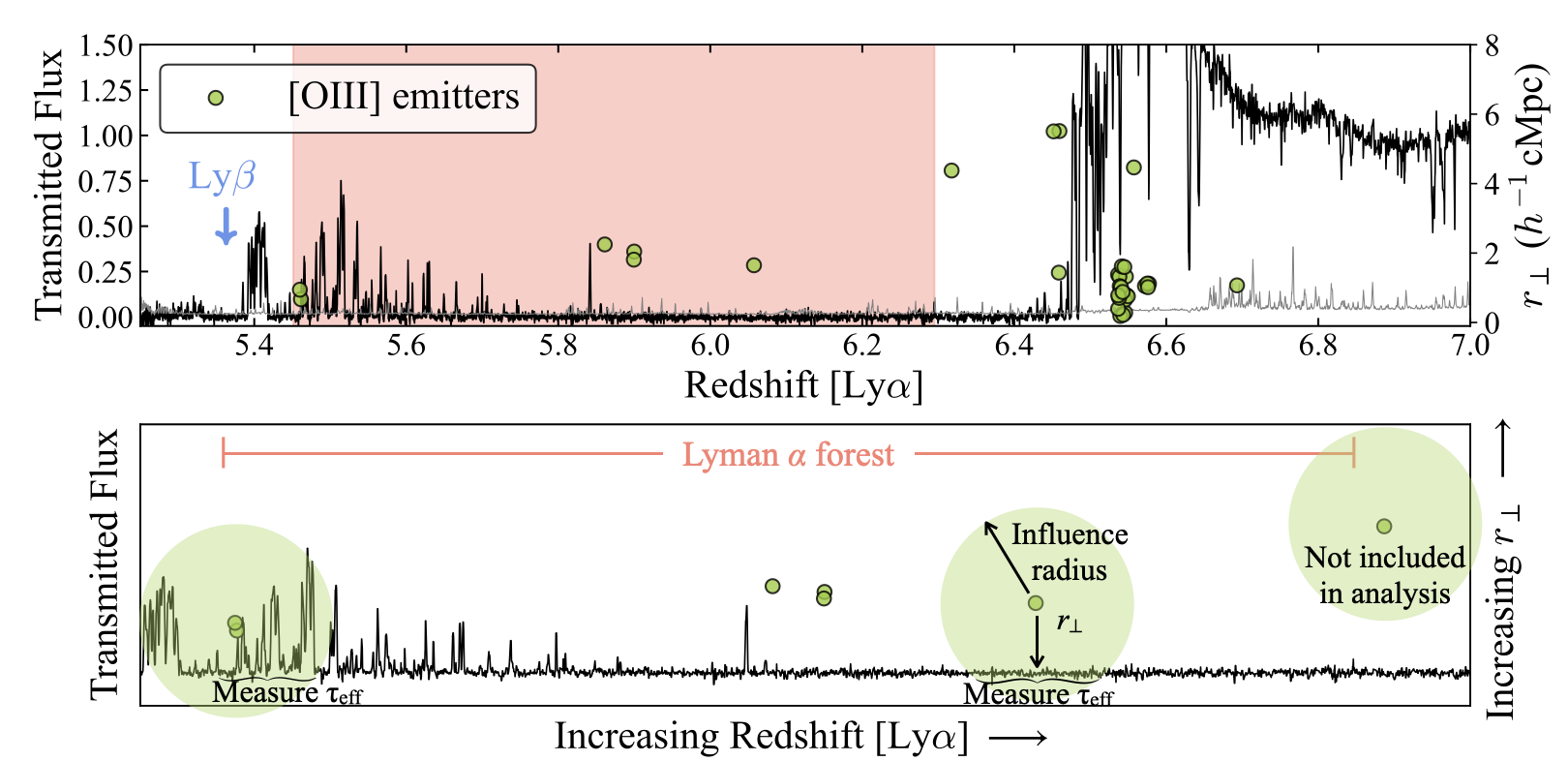}
    \caption{\textit{Top} -- The transmitted spectrum of an ASPIRE quasar J0226$+$0302. The redshift range of the Ly$\alpha$ used in the analysis is displayed in the red shaded region. The location of Ly$\beta$ emission line is marked by the blue downwards arrow. The spatial locations of the \oiii\ emitters identified in each quasar field are denoted by the yellowgreen circles, in terms of the \oiii\ emitter redshift $z_{\rm [OIII]}$ and the transverse distance $r_{\rm \perp}$ between the \oiii\ emitter and the central quasar. \textit{Bottom} -- An illustrative figure for measuring the IGM transmission around the \oiii\ emitters within the Ly$\alpha$ forest of J0226$+$0302. For each \oiii\ emitter identified in the quasar field, we put a sphere of a certain radius (``influence radius") centered at the \oiii\ emitter. For better visualization, we only show the spheres of three \oiii\ emitters and the size of the sphere is only for illustrative purposes, not to scale. For an \oiii\ emitter of which the sphere intersects the Ly$\alpha$ forest, we assume that the \oiii\ emitter is able to ``influence" the Ly$\alpha$ forest. 
    %If the sphere intersects the quasar sightline (i.\,e.,\, $r_{\perp}<$ influence radius), we measure the IGM effective optical depth over the path length enclosed by the sphere around the \oiii\ emitter. If the sphere centered at an \oiii\ emitter does not intersect the quasar sightline at all, we regard the \oiii\ emitter is too distant to influence the Ly$\alpha$ forest, and thus we do not include them in the analysis. For some \oiii\ emitters located close to the boundary of the Ly$\alpha$ forest, we measure the IGM effective optical depth only over the path length within the Ly$\alpha$ forest.
    }
    \label{fig:illustrative}
\end{figure*}

%{\color{red} One sentence to describe that the spectral resolution varies and the depth varies; Better to measure in a large bin size instead in individual pixel scales; %Or in the instroduction}
\subsection{IGM Ly$\alpha$ $\tau_{\rm eff}$ Measurements}\label{sec:IGMtaueff}

Because the spectral resolution and the exposure time of optical spectroscopy varies, we use a large bin size to measure the IGM transmission through $\tau_{\rm eff}$ measurements. 
% instead of measuring the IGM transmission on the original pixel scales. 
We first measure $\tau_{\rm eff}$ of each individual quasar sightline, defined as $\tau_{\rm eff}=-{\rm ln}\langle T_{\rm Ly\alpha}\rangle$, where $T_{\rm Ly\alpha}$ is the transmitted flux in the Ly$\alpha$ forest mentioned in Section \ref{sec:data}. We follow \citet{Becker2015MNRAS}, \citet{Bosman2018MNRAS,Bosman2022MNRAS}, \citet{Eilers2018ApJ} and \citet{Yang2020ApJb}, adopting a bin size of 50$\;\!h^{-1}$~cMpc, and measure $\tau_{\rm eff}$ in 50$\;\!h^{-1}$~cMpc bins, starting from the rest frame 1040~\AA\ up to the rest frame 1176~${\rm \AA}$. If the transmitted flux is not detected in a 50$\;\!h^{-1}$~cMpc bin (i.\,e.,\, $\langle T_{\rm Ly\alpha} \rangle < \langle 2\sigma/F_{\rm continuum} \rangle$), in this case, we use $\tau_{\rm lim, 2\sigma}$ as the $2\sigma$ lower limit of $\tau_{\rm eff}$. 

To examine the overall $\tau_{\rm eff}$ distribution of 14 ASPIRE quasar sightlines, we stack their Ly$\alpha$ forest spectrum. As the spectral resolution of different instruments varies, we use a common redshift grid with ${\rm d}z=0.001$ to stack the Ly$\alpha$ forest spectrum, 
adopting inverse variance weighting to avoid contamination from sky emission line residuals. The spectral uncertainty of the stacked spectrum is also calculated using the inverse variance weighting. The inverse variance used for weighting has been normalized by the quasar continuum flux at the corresponding rest wavelength.
%adopting inverse variance weighting to avoid contamination from sky emission line residuals. The inverse variance used for weighting has also been normalized by the quasar continuum flux at the corresponding rest wavelength. 
The stacked Ly$\alpha$ forest spectrum is shown in the top panel of Figure \ref{fig:stacked_lya_forest_spectrum}. The bottom panel of Figure \ref{fig:stacked_lya_forest_spectrum} shows the number of quasar sightlines available at each redshift grid. Nearly all quasar sightlines used in this work cover a Ly$\alpha$ absorption redshift range of $z\sim5.70-6.27$. We then measure the $\tau_{\rm eff}$ of the stacked Ly$\alpha$ forest spectrum in 50$\;\!h^{-1}$~cMpc bins, starting from $z\sim5.47$ up to $z\sim6.48$. 

Figure \ref{fig:stack_tau_eff} shows $\tau_{\rm eff}$ of 14 individual quasar sightlines and $\tau_{\rm eff}$ of the stacked Ly$\alpha$ forest spectrum (i.\,e.,\,Figure \ref{fig:stacked_lya_forest_spectrum}). The $\tau_{\rm eff}$ measurements for the single sightlines show a rapid increase and a large scatter as the trend observed in the previous works at $z>5.5$ \citep{Fan2006AJ,Becker2015MNRAS,Bosman2018MNRAS,Bosman2022MNRAS,Eilers2018ApJ,Yang2020ApJb}. 
%There are no IGM Ly$\alpha$ effective optical depth measurements at $z<5.4$ to exclude from contamination from \ion{O}{6}$\lambda$~1033 emission and the Ly$\beta$ forest. 
We show the best-fit $\tau_{\rm eff}$ redshift evolution from \citet{Yang2020ApJb} and \citet{Bosman2022MNRAS}. The best-fit $\tau_{\rm eff}$ redshift evolution from \citet{Yang2020ApJb} displays a higher $\tau_{\rm eff}$ than \citet{Bosman2022MNRAS} at the same redshift. \citet{Bosman2022MNRAS} argues that this difference could be attributed to cosmic variance of the different quasar samples as well as to the different quasar continuum reconstruction methods. \citet{Yang2020ApJb} adopt a broken power-law to calculate the quasar continuum flux, while \citet{Bosman2022MNRAS} use principal component analysis (PCA). \citet{Bosman2021MNRAS} suggest the power-law continuum reconstruction has a mean bias of 9.58\% over the Ly$\alpha$ forest because it does not fit broad emission lines, but it can accurately recover the continuum emission between broad emission lines. In this work, we analyze the quasar rest-frame wavelength range 1040--1176\AA\, which does not include broad emission lines such as Ly$\beta$ and \ion{O}{6}. Therefore, the power-law continuum should not introduce significant biases to the $\tau_{\rm eff}$ calculation.
%Therefore we directly compare the $\tau_{\rm eff}$ distribution with the best-fit $\tau_{\rm eff}$ redshift evolution in \citet{Yang2020ApJb}, to avoid any \textcolor{blue}{\textbf{potential}} systematic difference caused by different continuum reconstruction methods. 
The $\tau_{\rm eff}$ of the stacked Ly$\alpha$ forest transmission of 14 ASPIRE quasars falls between the best-fit $\tau_{\rm eff}$ redshift evolution derived from \citet{Yang2020ApJb} and \citet{Bosman2022MNRAS} at $z<6.1$, and at $z>6.1$, the stacked $\tau_{\rm eff}$ are lower limits. The $\tau_{\rm eff}$ distribution shown in Figure \ref{fig:stack_tau_eff} provides a baseline for IGM Ly$\alpha$ transmission, against which the IGM Ly$\alpha$ transmission around \oiii\ emitters will be evaluated.

\subsection{Measuring the IGM Ly$\alpha$ Transmission around \oiii\ Emitters}

With \oiii\ emitters selected by JWST NIRCam WFSS observations, it is now possible to study the connection between IGM transmission and galaxies. The top panel of Figure \ref{fig:illustrative} shows the transmitted spectrum of an ASPIRE quasar J0226$+$0302, together with the spatial location of all \oiii\ emitters identified in the same quasar field. The transmitted spectrum of all 14 ASPIRE quasars used in this work can be found in Appendix \ref{appendix:los}. The spatial locations of \oiii\ emitters are denoted by the \oiii\ redshift $z_{\rm [OIII]}$ of the identified \oiii\ emitter and the transverse distance $r_{\rm \perp}$ between the \oiii\ emitter and the quasar. 

To probe the connection between IGM transmission and galaxies, we put a sphere centered at the detected \oiii\ emitters with a certain radius (hereafter ``influence radius"). If the sphere intersects the Ly$\alpha$ forest, we then measure $\tau_{\rm eff}$ over the path length enclosed by the sphere (the bottom panel of Figure \ref{fig:illustrative}). For \oiii\ emitters whose sphere does not intersect the Ly$\alpha$ forest, we then regard that those \oiii\ emitters are too distant to influence the quasar Ly$\alpha$ forest. 
%therefore we exclude them in our analysis. 
For \oiii\ emitters close to the boundary of the Ly$\alpha$ forest included in our analysis, we only use the path length enclosed by the sphere and within the Ly$\alpha$ forest to measure $\tau_{\rm eff}$. 
%Because the FoV of JWST NIRCam is small, 
%($\sim5.1\;\!$cMpc$\times5.1\;\!$cMpc for each module at $z\sim5.5$), 
%only with small influence radii ($\lesssim7~h^{-1}\;\!$cMpc), some \oiii\ emitters will be excluded from our analysis. 
%When the influence radius is greater than $7~h^{-1}\;\!$cMpc, all \oiii\ emitters within the redshift range of the Ly$\alpha$ forest are included in our analysis. 
By measuring $\tau_{\rm eff}$ around \oiii\ emitters, we sample the IGM transmission around them. As such, the resulting $\tau_{\rm eff}$ distribution will be a biased distribution including the galaxy proximity effect.

\begin{figure*}[!ht]
    \centering
    \includegraphics[width=0.9\textwidth]{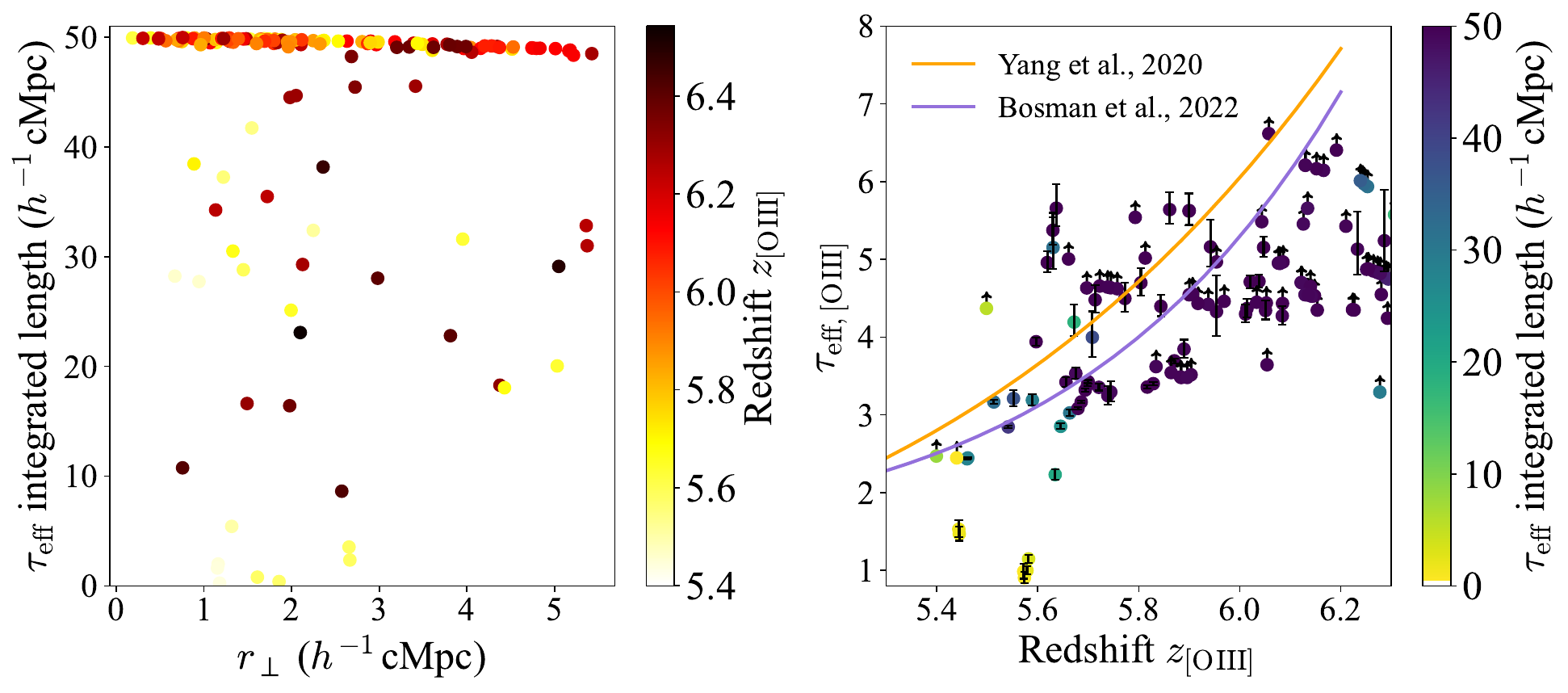}
    \caption{\textit{Left} -- The integrated path length for $\tau_{\rm eff, [O III]}$ measurements as a function of the transverse distance ($r_{\perp}$), adopting an influence radius of 25~$h^{-1}\;\!{\rm cMpc}$. Each point is color-coded by the redshift of the \oiii\ emitter $z_{\rm [O III]}$. Except for those \oiii\ emitters at the boundary of the Ly$\alpha$ forest, for most \oiii\ emitters, the integrated length for $\tau_{\rm eff, [O III]}$ measurements is $\sim50~h^{-1}\;\!{\rm cMpc}$. \textit{Right} -- The $\tau_{\rm eff}$ around the \oiii\ emitters ($\tau_{\rm eff, [O III]}$) adopting an influence radius of 25~$h^{-1}\;\!{\rm cMpc}$, as the function of the \oiii\ emitter redshift $z_{\rm [O III]}$. The integrated path length for each $\tau_{\rm eff, [O III]}$ measurement is color-coded. The best-fit $\tau_{\rm eff}$ redshift evolution from \citet{Yang2020ApJb} and \citet{Bosman2022MNRAS} is denoted by the orange and purple line. }
    \label{fig:IGM_tau_eff_o3e}
\end{figure*}

%{\color{red}{add justification of the selection of influence radius}} 
We first adopt an influence radius of $25~h^{-1}\;\!$cMpc and then measure the $\tau_{\rm eff}$ around \oiii\ emitters (hereafter $\tau_{\rm eff,[OIII]}$). In this case, twice the influence radius is equal to $50~h^{-1}\;\!$cMpc, which is the bin size used for $\tau_{\rm eff}$ measurements in Figure \ref{fig:stack_tau_eff}. 
%In this case, the integrated path length for $\tau_{\rm eff,[OIII]}$ measurements is roughly twice the influence radius ($\sim50~h^{-1}\;\!$cMpc), comparable to the bin size used for $\tau_{\rm eff}$ measurements shown in Figure \ref{fig:stack_tau_eff}. 
We show the dependence of $\tau_{\rm eff,[OIII]}$ integrated length on the transverse distance $r_{\perp}$ in the left panel of Figure \ref{fig:IGM_tau_eff_o3e}. 
%Because we adopt an influence radius of $25~h^{-1}\;\!$cMpc, significantly larger than the maximal transverse distance of $\sim5~h^{-1}\;\!$cMpc between \oiii\ emitters and the quasar, 
Because all detected \oiii\ emitters in ASPIRE fields have $r_{\perp}<<50~h^{-1}\;\!$cMpc, for most $\tau_{\rm eff,[OIII]}$ measurements, by adopting an influence radius of $25~h^{-1}\;\!$cMpc, the corresponding integrated path length is $\sim49-50~h^{-1}\;\!$cMpc, 
%very close to the 50~$h^{-1}\;\!{\rm cMpc}$ used in the IGM Ly$\alpha$ effective optical depth measurements in Figure \ref{fig:stack_tau_eff}, 
except for those \oiii\ emitters close to the boundary of the Ly$\alpha$ forest (i.e., $z_{\rm [OIII]}<5.7$ or $z_{\rm [OIII]}>6.3$). This ensures that most $\tau_{\rm eff,[OIII]}$ is measured with a nearly uniform integrated length of $\sim50~h^{-1}\;\!$cMpc as $\tau_{\rm eff}$ measurements in Figure \ref{fig:stack_tau_eff}. We show $\tau_{\rm eff,[OIII]}$ as a function of the \oiii\ emitter redshift $z_{\rm [OIII]}$ in the right panel of Figure \ref{fig:IGM_tau_eff_o3e}. If the IGM Ly$\alpha$ transmission is not detected ($<2\sigma$) around an \oiii\ emitter, we then use $\tau_{\rm lim, 2\sigma}$ as the $2\sigma$ lower limit of $\tau_{\rm eff,[OIII]}$. Similar to the redshift evolution of $\tau_{\rm eff}$ shown in Figure \ref{fig:stack_tau_eff}, $\tau_{\rm eff,[OIII]}$ around individual \oiii\ emitters also shows an increasing trend as the redshift increases and displays a large scatter. Because the integrated length is shorter for $z\lesssim5.7$ $\tau_{\rm eff,[OIII]}$ measurements, the scatter in $\tau_{\rm eff,[OIII]}$ is larger at $z\lesssim5.7$ than at $z\gtrsim5.7$, reflecting small scale variations in IGM transmission. At $z\gtrsim5.7$, the redshift evolution of $\tau_{\rm eff,[OIII]}$ displays a slightly flatter trend than the redshift evolution of $\tau_{\rm eff}$ denoted by the orange line in the right panel of Figure \ref{fig:IGM_tau_eff_o3e}. At $z>6$, most $\tau_{\rm eff,[OIII]}$ measurements are lower limits due to the depth of the current data.  %The overall trend is systematically lower than the average IGM Ly$\alpha$ $\tau_{\rm eff}$ redshift evolution measured from \citet{Yang2020ApJ}. 

%The difference between $\tau_{\rm eff,[OIII]}$ and $\tau_{\rm eff}$ is not subjective to the difference in the integrated length. {\color{red}{re-word?}}

%\section{Results and Discussion} \label{sec:result}
\section{IGM Patches Are More Transparent around \oiii\ Emitters} \label{sec:result}

\begin{comment}
\begin{itemize}
    \item Measure the IGM tau around \oiii\ emitters. -- summarize the main point from the last session; describe how we measure it
    \item Mention the integrate length
    \item CDF and describe of random CDF
    \item results with different influence radii 
    \item discussion: implication for reionization topology; size of the ionizing bubbles? 
\end{itemize}
\end{comment}

%\subsection{Stacked Ly$\alpha$ Forest Spectrum around \oiii\ Emitters}

%From CDFs of IGM effective optical depth around $5.4<z<6.1$ \oiii\ emitters, there is an excess of IGM transmission when adopting an influence radius of $25~h^{-1}\;\!$cMpc. 

\begin{figure*}[!ht]
    \centering
    \includegraphics[width=0.95\textwidth]{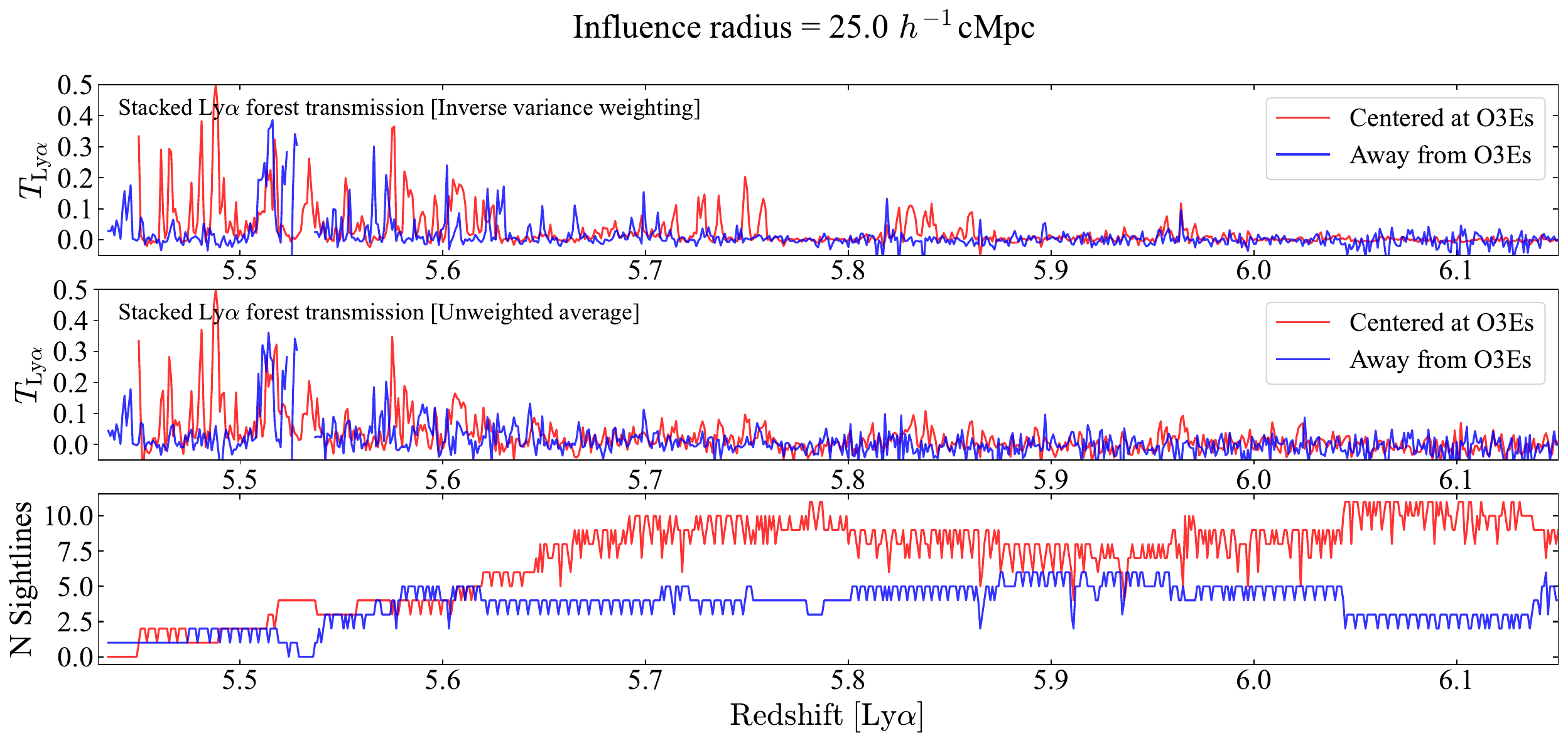}
    \caption{\textit{Top} - The stacked transmission of the Ly$\alpha$ forest around \oiii\ emitters (red) and away from \oiii\ emitters (blue), adopting the inverse variance weighting and an influence radius of 25~$h^{-1}\;\!{\rm cMpc}$, see text for details. \textit{Middle} - The stacked transmission of the Ly$\alpha$ forest around \oiii\ emitters (red) and away from \oiii\ emitters (blue), where the stacked transmission is the unweighted average transmission. \textit{Bottom} - The number of quasar line of sights (LOSs) available in each pixel of the stacked Ly$\alpha$ forest transmission around \oiii\ emitters (red) and the stacked Ly$\alpha$ forest transmission away from \oiii\ emitters (blue).}
    \label{fig:stacked_spectrum_wo3e}
\end{figure*}

We first investigate $\tau_{\rm eff}$ of the stacked Ly$\alpha$ forest spectrum centered at \oiii\ emitters using an influence radius of $25~h^{-1}\;\!$cMpc used in Figure \ref{fig:IGM_tau_eff_o3e}. %and the stacked Ly$\alpha$ forest spectrum where no \oiii\ emitters are detected. 
By adopting a certain influence radius, for the Ly$\alpha$ forest spectrum of each quasar sightline, we split the Ly$\alpha$ forest spectrum into two parts: (1) the Ly$\alpha$ forest spectrum centered at \oiii\ emitters (i.\,e.,\,the part of Ly$\alpha$ forest spectrum that intersects the sphere of \oiii\ emitters); and (2) the Ly$\alpha$ forest spectrum away from \oiii\ emitters, where no \oiii\ emitters are detected (i.\,e.,\,the part of Ly$\alpha$ forest spectrum that does not intersect the sphere of any detected \oiii\ emitters). We then stack the Ly$\alpha$ forest spectrum centered at \oiii\ emitters and the Ly$\alpha$ forest spectrum away from \oiii\ emitters separately, using the inverse variance weighting and the same redshift grid as Figure \ref{fig:stacked_lya_forest_spectrum}. The top panel of Figure \ref{fig:stacked_spectrum_wo3e} shows the stacked spectrum of the Ly$\alpha$ forest spectrum around \oiii\ emitters and away from \oiii\ emitters, and the bottom panel of Figure \ref{fig:stacked_spectrum_wo3e} shows the number of quasar sightlines in each pixel. 

By comparing the stacked entire Ly$\alpha$ forest spectrum (Figure \ref{fig:stacked_lya_forest_spectrum}) and stacked Ly$\alpha$ forest spectrum centered at \oiii\ emitters (the red line in the top panel of Figure \ref{fig:stacked_spectrum_wo3e}), the majority of the Ly$\alpha$ transmission in the stacked entire Ly$\alpha$ forest spectrum is around \oiii\ emitters (within an influence radius of $25~h^{-1}\;\!$cMpc). 
Compared with the stacked Ly$\alpha$ forest spectrum away from \oiii\ emitters, the stacked Ly$\alpha$ forest spectrum centered at \oiii\ emitters displays more prominent IGM transmission at $z>5.7$. 
To test whether the result is mainly caused by the bias towards high S/N quasar sightlines, we generate the stacked Ly$\alpha$ transmission spectrum using the unweighted average transmission. 
The unweighted average Ly$\alpha$ transmission centered at \oiii\ emitters and away from \oiii\ emitters is shown in the middle panel of Figure \ref{fig:stacked_spectrum_wo3e}. Compared with the stacked Ly$\alpha$ transmission calculated by the inverse variance weighting, the unweighted average Ly$\alpha$ transmission shows more pixel-to-pixel variations, but a smoother evolution in redshift. The unweighted average Ly$\alpha$ transmission around \oiii\ emitters is still more prominent than the unweighted average Ly$\alpha$ transmission away from \oiii\ emitters at the same redshift. 
%By adopting different weightings, the corresponding stacked Ly$\alpha$ transmission around \oiii\ emitters is still more prominent than the Ly$\alpha$ transmission away from \oiii\ emitters at the same redshift. 
This suggests that regions traced by \oiii\ emitters are playing important roles in contributing local ionizing background and producing the observed transmission in the Ly$\alpha$ forest. 
%This is consistent with simulated results in \citet{Garaldi2019ApJ}, where most transmission spikes are in the vicinity of UV bright sources. 

\begin{figure}[!ht]
    \centering
    \includegraphics[width=0.5\textwidth]{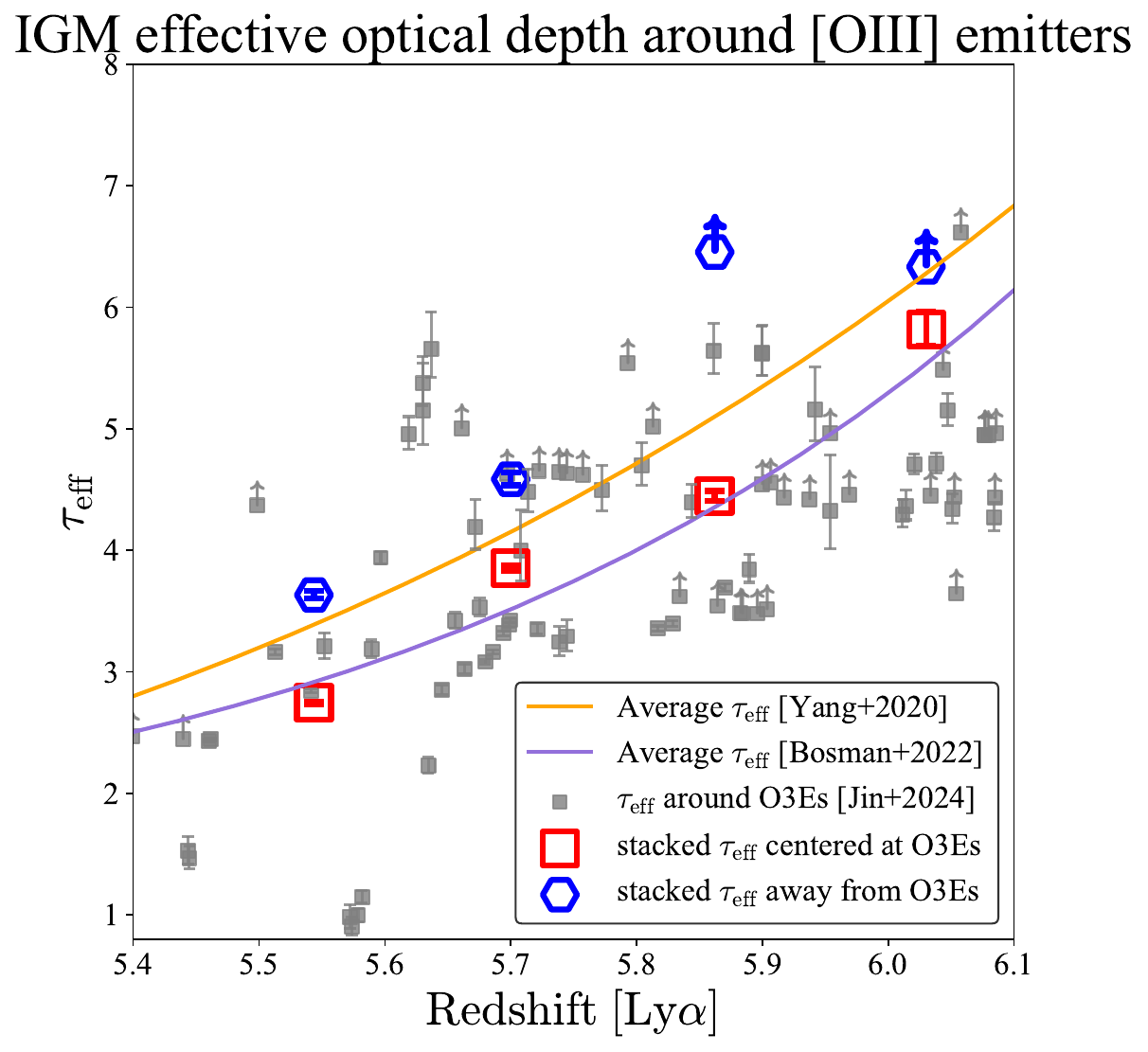}
    \caption{The $\tau_{\rm eff}$ measurements of the stacked Ly$\alpha$ forest spectrum centered at \oiii\ emitters (the red line in the top panel of Figure \ref{fig:stacked_spectrum_wo3e}) are denoted by red open squares, and the $\tau_{\rm eff}$ measurements of the stacked Ly$\alpha$ forest spectrum away from \oiii\ emitters (the blue line in the top panel of Figure \ref{fig:stacked_spectrum_wo3e}) are marked by blue open hexagons. Individual measurements of IGM optical depth around \oiii\ emitters $\tau_{\rm eff,[OIII]}$ are shown in filled grey squares, adopting an influence radius of 25~$h^{-1}\;\!{\rm cMpc}$. The best-fit redshift evolutions of $\tau_{\rm eff}$ from \citet{Yang2020ApJb} and \citet{Bosman2022MNRAS} are denoted by the orange and the purple line.}
    \label{fig:IGM_tau_eff_stacked}
\end{figure}

We then measure the $\tau_{\rm eff}$ of stacked Ly$\alpha$ transmission spectra centered at and away from \oiii\ emitters (the top panel of Figure \ref{fig:stacked_spectrum_wo3e}). We show the $\tau_{\rm eff}$ measurements in Figure \ref{fig:IGM_tau_eff_stacked} by the red hollow square markers (centered at \oiii\ emitters) and the blue hollow hexagon markers (away from \oiii\ emitters). The $\tau_{\rm eff}$ of the stacked Ly$\alpha$ forest spectrum centered at \oiii\ emitters is significantly ($>5\sigma$) lower than the $\tau_{\rm eff}$ of the stacked Ly$\alpha$ forest spectrum away from \oiii\ emitters. At the same redshift, the IGM patches away from \oiii\ emitters have an average $\tau_{\rm eff}$ of $\gtrsim0.5$ higher than the $\tau_{\rm eff}$ of the IGM patches around \oiii\ emitters, suggesting that near \oiii\ emitters, the IGM patches are more transparent than those regions in the IGM where no \oiii\ emitters are detected. This indicates a higher IGM transmission near \oiii\ emitters and suggests that \oiii\ emitters can enhance the ionizing radiation field around them, resulting in a higher local ionizing background. This implies the scatter in the observed $\tau_{\rm eff}$ at $z>5.5$ is tightly associated with the fluctuations in the ionizing background \citep{Fan2006AJ,Davies2024ApJ}. 
By comparing the same optical depth around \oiii\ emitters and away from \oiii\ emitters, we find the IGM patches around \oiii\ emitters reach the same $\tau_{\rm eff}$ around ${\rm d}z\sim0.1$ earlier than IGM patches where no \oiii\ emitters are detected, suggesting reionization processes faster ($\gtrsim$23~Myr ahead) around \oiii\ emitters before the end of reionization \cite[$z\sim5.3$, ][]{Bosman2022MNRAS}.

\begin{figure*}[!ht]
    \centering
    \includegraphics[width=0.9\textwidth]{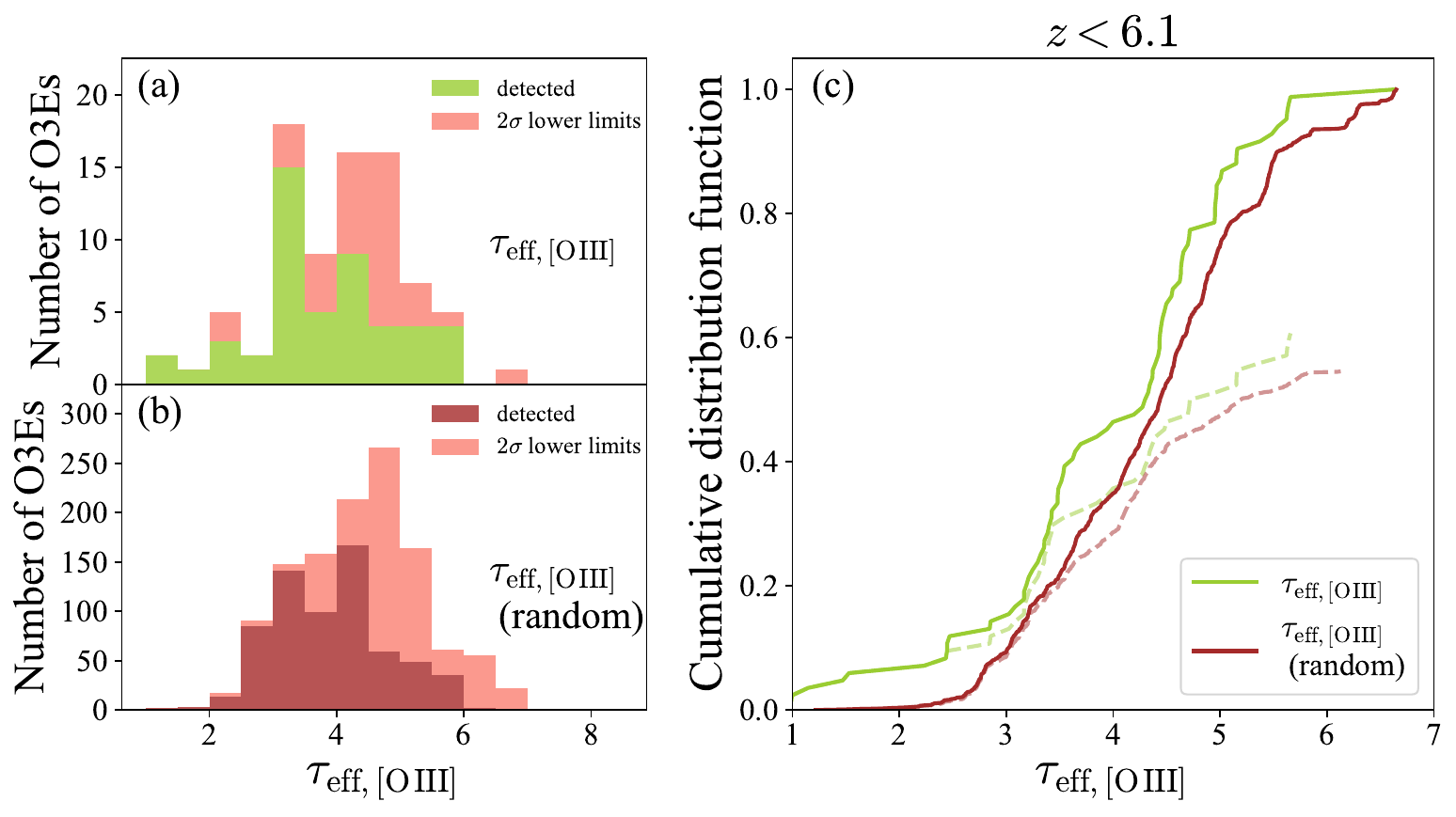}
    \caption{(a) The distribution of $\tau_{\rm eff}$ around the \oiii\ emitters ($\tau_{\rm eff,[OIII]}$) by adopting an influence radius of 25~$h^{-1}\;\!{\rm cMpc}$. (b) The random distribution of $\tau_{\rm eff,[OIII]}$ by adopting an influence radius of 25~$h^{-1}\;\!{\rm cMpc}$. The distributions of $2\sigma$ lower limits of $\tau_{\rm eff}$ are shown in red in (a) and (b). 
    %and the original distribution of IGM effective optical depth (bottom), 
    (c) The cumulative distribution functions (CDFs) of the $\tau_{\rm eff,[OIII]}$ (yellowgreen) and the random distribution (brown). For $2\sigma$ lower limits of $\tau_{\rm eff}$, we plot them as actual measurements (solid line) or infinity (dashed line) in CDF. %The $1\sigma$ uncertainty of the CDF derived from bootstrapping resampling is shown in the shaded regions of the corresponding color.
    }
    \label{fig:distribution_of_IGM_tau_eff_o3e}
\end{figure*} 

\begin{figure*}[htb!]
    \centering
    \includegraphics[width=1.0\textwidth]{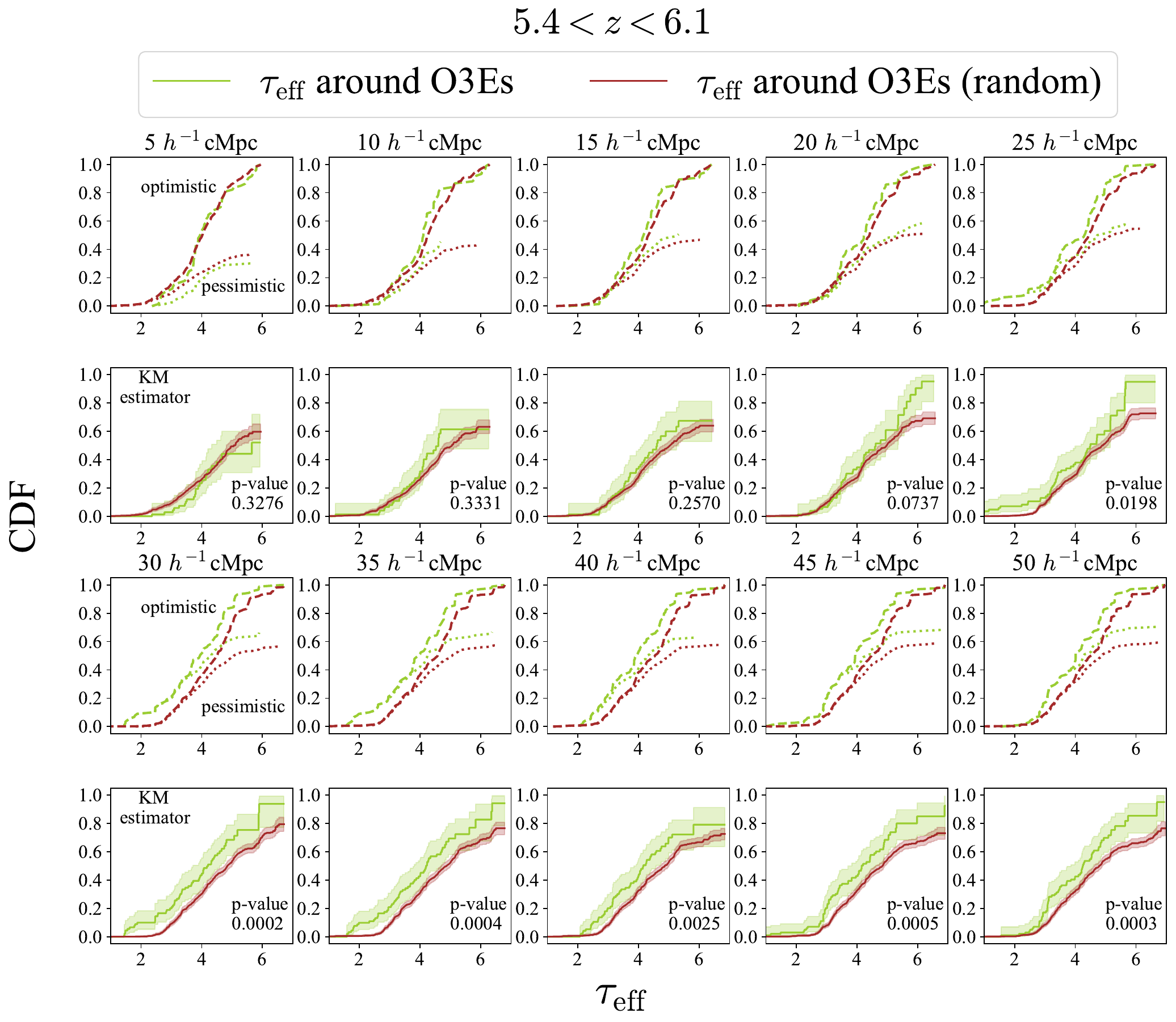}
    \caption{The first and the third rows show the cumulative distribution functions of the $\tau_{\rm eff}$ around \oiii\ emitters at $5.4<z<6.1$ (yellowgreen), adopting influence radii from 5~$h^{-1}\;\!{\rm cMpc}$ to 50~$h^{-1}\;\!{\rm cMpc}$. The random cumulative distribution function of the $\tau_{\rm eff}$ around \oiii\ emitters is shown in brown, representing the $\tau_{\rm eff}$ distribution of IGM patches. The ``optimistic" and the ``pessimistic" CDFs are plotted in dashed and dotted lines, respectively. The second and the fourth rows show the fitted CDFs from Kaplan-Meier (KM) estimator and $1\sigma$ confidence interval of the CDF is shown in the shaded regions of the corresponding color. The null-hypothesis p-value from the log-rank test is shown in the bottom right corner of each sub-panel. See text for more details. %With an influence radius greater than 30~$h^{-1}\;\!{\rm cMpc}$, the IGM effective optical depth around \oiii\ emitters is significantly different than the random distribution, indicating that the IGM effective optical depth around \oiii\ emitters is significantly lower.
    }
    \label{fig:cdf_different scales}
\end{figure*}

\begin{figure*}
    \centering
    \includegraphics[width=1.0\textwidth]{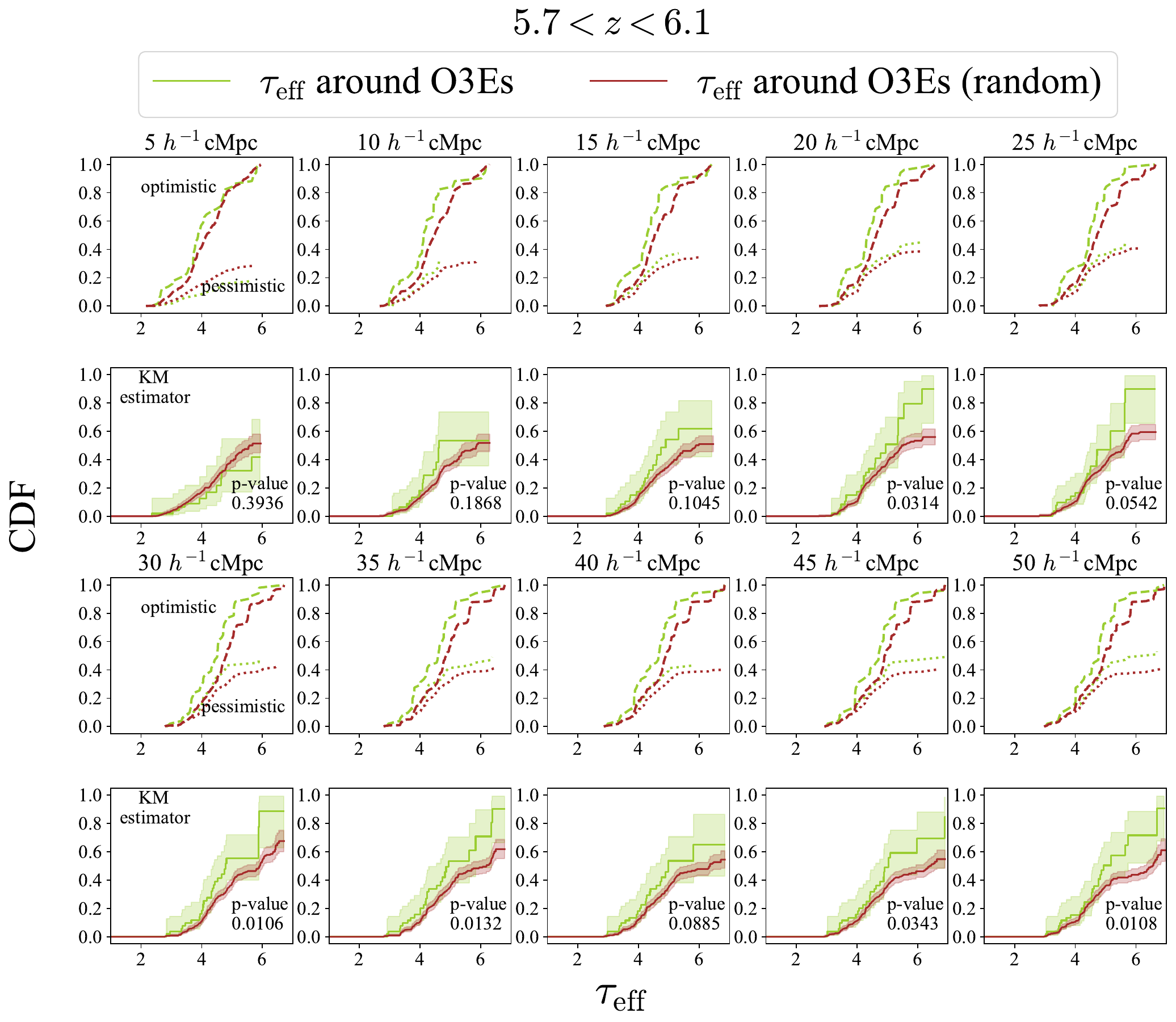}
    \caption{Similar to Figure \ref{fig:cdf_different scales}, but for \oiii\ emitters at $5.7<z<6.1$.}
    \label{fig:cdf_different scales_high_z}
\end{figure*}

\begin{figure*}
    \centering
    \includegraphics[width=1.0\textwidth]{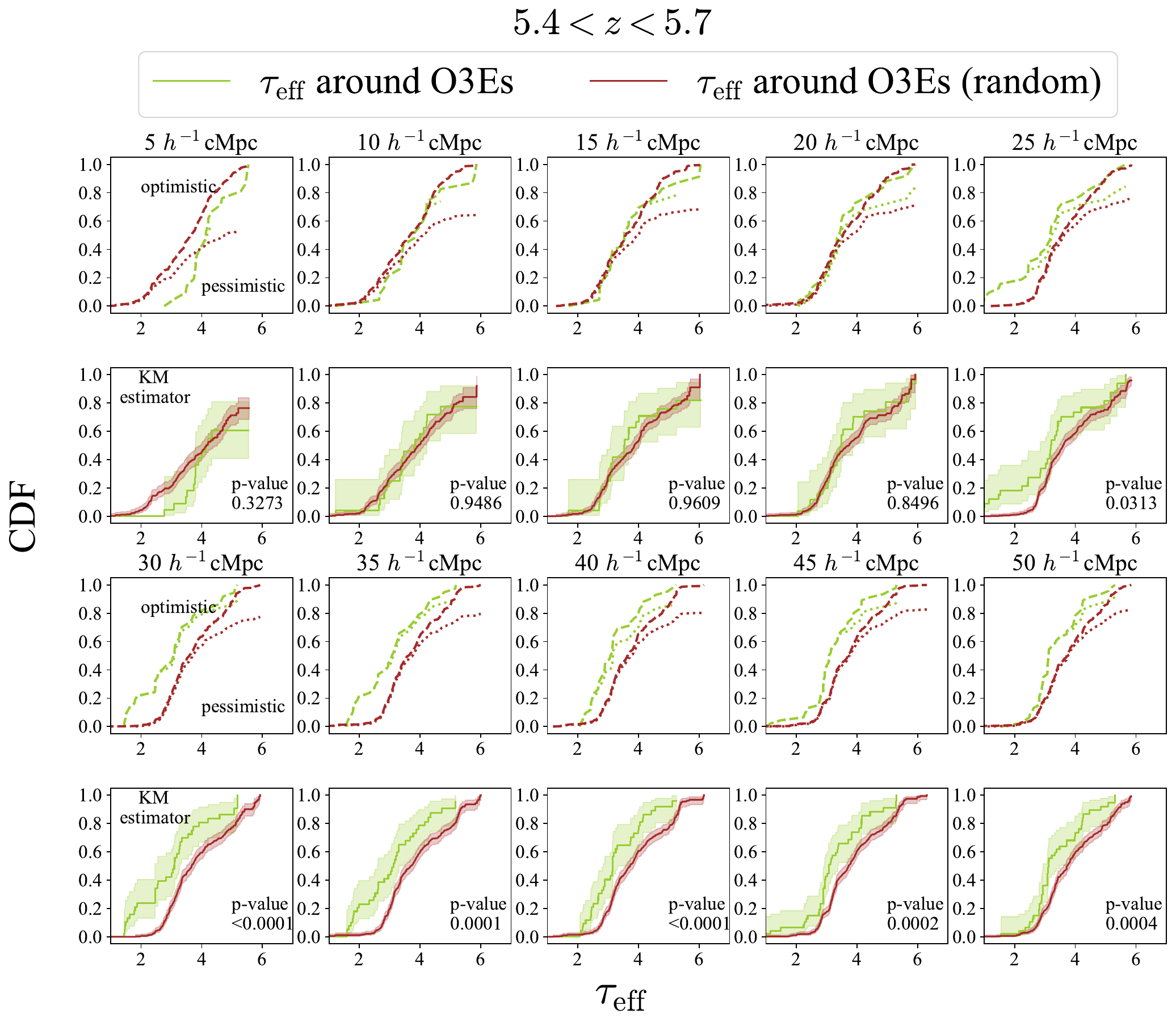}
    %\caption{The cumulative distribution functions of the IGM effective optical depth around \oiii\ emitters at $5.4<z<5.7$ (yellowgreen) and the random distribution (brown), adopting influence radii from 5~$h^{-1}\;\!{\rm cMpc}$ to 50~$h^{-1}\;\!{\rm cMpc}$. The $1\sigma$ uncertainty of the CDF derived from bootstrapping sampling is shown in the shaded regions of the corresponding color. The null-hypothesis p-value for AD tests is shown in the upper left corner of each subpanel. With an influence radius of 5~$h^{-1}\;\!{\rm cMpc}$, the IGM effective optical depth around \oiii\ emitters is significantly higher than the random distribution, suggesting Ly$\alpha$ absorption closed to \oiii\ emitters due to gas overdensity. Significant excess of IGM transmission around \oiii\ emitters is shown beyond an influence radius of 25~$h^{-1}\;\!{\rm cMpc}$.}
    \caption{Similar to Figure \ref{fig:cdf_different scales_high_z}, but for \oiii\ emitters at $5.4<z<5.7$}
    \label{fig:cdf_different scales_low_z}
\end{figure*}

\section{Discussion}\label{sec:discussion}

In this section, we examine the distribution of $\tau_{\rm eff}$ around \oiii\ emitters (i.\,e.,\,$\tau_{\rm eff,[OIII]}$) at different redshifts and on various scales, and investigate whether IGM patches around \oiii\ emitters are biased towards more transparent or more opaque IGM patches, thereby constraining the topology of reionization. 

\subsection{IGM Effective Optical Depth Distribution around $z<6.1$ \oiii\ Emitters}\label{sec:discussion_CDF_blz6p1}

Although all ASPIRE quasars used in this work cover a Ly$\alpha$ absorption redshift at $z>6.1$, given the depth of the current data, $>75\%$ of measurements are $2\sigma$ lower limits if we include $z>6.1$ $\tau_{\rm eff}$ measurements (see Figure \ref{fig:stack_tau_eff}). Therefore, we only investigate the distribution of $\tau_{\rm eff,[OIII]}$ at $z<6.1$.

By adopting an influence radius of $25~h^{-1}\;\!$cMpc, most integrated lengths are close to the bin size ($50~h^{-1}\;\!$cMpc) used in previous IGM $\tau_{\rm eff}$ studies (see \citet{Becker2015MNRAS}, \citet{Bosman2018MNRAS}, \citet{Eilers2018ApJ} and \citet{Yang2020ApJb}). We show the distribution of $\tau_{\rm eff,[OIII]}$ around $z<6.1$ \oiii\ emitters in a histogram in Figure \ref{fig:distribution_of_IGM_tau_eff_o3e}(a). To compare the $\tau_{\rm eff,[OIII]}$ distribution with the $\tau_{\rm eff}$ distribution measured in Section \ref{sec:IGMtaueff}, we generate a random $\tau_{\rm eff,[OIII]}$ distribution. We utilize the spatial location of $z<6.1$ \oiii\ emitters detected in the other 13 quasar fields and re-compute $\tau_{\rm eff,[OIII]}$ for each quasar field, based on the methods described in Section \ref{sec:method}. We show the random $\tau_{\rm eff,[OIII]}$ distribution in a brown line in Figure \ref{fig:distribution_of_IGM_tau_eff_o3e}(b). The random $\tau_{\rm eff,[OIII]}$ distribution provides a control sample to compare with the $\tau_{\rm eff,[OIII]}$ distribution. Ideally, the random distribution can well represent the original $\tau_{\rm eff}$ distribution of IGM patches measured in Section \ref{sec:IGMtaueff}. We compare the random $\tau_{\rm eff,[OIII]}$ distribution with the original $\tau_{\rm eff}$ distribution of IGM patches in Appendix \ref{appendix:random_check}, and find the random $\tau_{\rm eff,[OIII]}$ distribution is consistent with the original $\tau_{\rm eff}$ distribution of IGM patches. 

In Figure \ref{fig:distribution_of_IGM_tau_eff_o3e}(a) and Figure \ref{fig:distribution_of_IGM_tau_eff_o3e}(b), the distribution of $2\sigma$ lower limits of $\tau_{\rm eff,[OIII]}$ is shown in red. The distribution of $\tau_{\rm eff,[OIII]}$ displays a more significant peak at a $\tau_{\rm eff}$ of $\sim3$ than the random distribution. In Figure \ref{fig:distribution_of_IGM_tau_eff_o3e}(c), we show the cumulative distribution functions (CDFs) of the $\tau_{\rm eff,[OIII]}$ distribution and the random $\tau_{\rm eff,[OIII]}$ distribution. Because there are lower limits in both distributions, following \citet{Bosman2018MNRAS,Bosman2022MNRAS}, we treat the lower limits as either (1) the ``optimistic" case: the transmitted flux is just below the 2$\sigma$ detection limit (i.e., $\tau_{\rm eff}=\tau_{\rm lim, 2\sigma}$) or (2) the ``pessimistic" case: the intrinsic transmitted flux is zero (i.e., $\tau_{\rm eff}=\infty$). The CDFs of these two different cases are plotted in solid lines ($\tau_{\rm eff}=\tau_{\rm lim, 2\sigma}$) and dashed lines ($\tau_{\rm eff}=\infty$) in Figure \ref{fig:distribution_of_IGM_tau_eff_o3e}(c), respectively. The ``optimistic" and ``pessimistic" cases set the upper and lower bounds of the CDFs. The distribution of $\tau_{\rm eff,[OIII]}$ displays a higher cumulative probability % at a lower $\tau_{\rm eff}\sim3-4$ 
than the random distribution of $\tau_{\rm eff,[OIII]}$ for both ``optimistic" and ``pessimistic" cases. 
%{\textcolor{violet}{The optical depth of mean transmitted flux around \oiii\ emitters and the optical depth of mean transmitted flux around random \oiii\ emitters are denoted by the yellowgreen and brown upside down triangles. The optical depth of the mean transmitted flux is lower around \oiii\ emitters than the random distribution.}} 

In previous studies, the ``optimistic" CDF is often adopted when examining whether two distributions are consistent \cite[e.\,g.,\,][]{Becker2015MNRAS,Yang2020ApJb,Bosman2022MNRAS}, especially when comparing with those CDFs derived from simulations assuming homogenous UVB models \cite[e.\,g.,\,][]{Bosman2022MNRAS}. Because there are a number of lower limits (i.\,e.,\, right-censored data) in the $\tau_{\rm eff,[OIII]}$ distribution, we use \texttt{lifelines} package \citep{Davidson-Pilon2019}, to perform the log-rank test between the $\tau_{\rm eff,[OIII]}$ distribution and the random distribution to examine whether they are drawn from the same parent distribution. The p-value for null-hypothesis is 0.0198, suggesting that the CDFs of $\tau_{\rm eff,[OIII]}$ and random $\tau_{\rm eff,[OIII]}$ are different ($>2\sigma$) when adopting an influence radius of $25~h^{-1}\;\!$cMpc.
%\textcolor{blue}{\textbf{Following \citet{Becker2015MNRAS}, \citet{Yang2020ApJb} and \citet{Bosman2022MNRAS}, we adopt the ``optimistic" CDF to examine whether two distributions are consistent. We perform an Anderson-Darling (AD) test between the $\tau_{\rm eff,[OIII]}$ distribution and random distribution, and the p-value for null-hypothesis is 0.0029, suggesting that the CDFs of $\tau_{\rm eff,[OIII]}$ and random $\tau_{\rm eff,[OIII]}$ are different ($>2\sigma$) when adopting an influence radius of $25~h^{-1}\;\!$cMpc.}}

Furthermore, we measure $\tau_{\rm eff,[OIII]}$ of all $z<6.1$ \oiii\ emitters and corresponding random $\tau_{\rm eff,[OIII]}$ distributions by adopting different influence radii. We then investigate whether these two distributions will be significantly different at certain scales. Figure \ref{fig:cdf_different scales} shows the ``optimistic" and ``pessimistic" CDFs of $\tau_{\rm eff,[OIII]}$ and the random distribution using influence radii from $5~h^{-1}\;\!$cMpc to $50~h^{-1}\;\!$cMpc to demonstrate two distinct methods for handling lower limits in the distribution function. When the dataset includes censored data (i.e., lower limits or upper limits), survival analysis can be used to reconstruct the distribution function \citep{FN1985ApJ}. For each distribution containing $2\sigma$ lower limits of $\tau_{\rm eff,[OIII]}$, we use the Kaplan-Meier (KM) estimator, included in \texttt{lifelines} package \citep{Davidson-Pilon2019}, to fit the survival function. The $1\sigma$ confidence interval from the KM estimator of the fitted CDF is plotted as the shaded region in Figure \ref{fig:cdf_different scales}. The null-hypothesis p-value from the log-rank test is shown in the bottom right corner of each sub-panel. With influence radii $\lesssim20~h^{-1}\;\!$cMpc, the CDF of $\tau_{\rm eff,[OIII]}$ distribution is consistent with the random $\tau_{\rm eff,[OIII]}$ distribution with a p-value $>0.05$. When the influence radius is $\gtrsim25~h^{-1}\;\!$cMpc, the CDF of $\tau_{\rm eff,[OIII]}$ is significantly higher than the random $\tau_{\rm eff,[OIII]}$ distribution at the same $\tau_{\rm eff,[OIII]}$, with a null-hypothesis p-value $<0.05$, indicating that around $z<6.1$ \oiii\ emitters, the IGM patches are more transparent around \oiii\ emitters than elsewhere in the IGM on scales greater than $25~h^{-1}\;\!$cMpc, consistent with the topology of reionization predicted by the fluctuating UVB models.

%\textcolor{blue}{\textbf{For each ``optimistic" CDF, following \citet{Eilers2018ApJ} and \citet{Yang2020ApJb}, we adopt the bootstrapping method to obtain 5000 samples and then calculate the $1\sigma$ uncertainty of the CDF. The $1\sigma$ confidence interval is plotted as the shaded region in Figure \ref{fig:cdf_different scales}.}} 
%The null-hypothesis p-value from AD tests is shown in the upper left corner of each sub-panel. With influence radii $\lesssim20~h^{-1}\;\!$cMpc, the CDF of $\tau_{\rm eff,[OIII]}$ distribution is consistent with the random $\tau_{\rm eff,[OIII]}$ distribution. When the influence radius is $\gtrsim25~h^{-1}\;\!$cMpc, the CDF of $\tau_{\rm eff,[OIII]}$ is significantly higher than the random $\tau_{\rm eff,[OIII]}$ distribution at the same $\tau_{\rm eff,[OIII]}$, with a null-hypothesis p-value $<0.05$, indicating that around $z<6.1$ \oiii\ emitters, the IGM patches are more transparent around \oiii\ emitters than elsewhere in the IGM on scales greater than $25~h^{-1}\;\!$cMpc, consistent with the topology of reionization predicted by the fluctuating UVB models.  

\subsection{Redshift Evolution of IGM Effective Optical Depth Distribution around \oiii\ Emitters}\label{sec:discussion_CDF_z_evolve}

To further investigate whether the $\tau_{\rm eff}$ distribution around \oiii\ emitters evolves with redshift, by adopting different influence radii, we study the CDFs of $\tau_{\rm eff,[OIII]}$ in two different redshift bins: $5.7<z<6.1$ and $5.4<z<5.7$ in Figure \ref{fig:cdf_different scales_high_z} and Figure \ref{fig:cdf_different scales_low_z}. The selection of the redshift bins is motivated by: (1) \citet{Kashino2023ApJ} shows the IGM-galaxy cross-correlation function signal can change with redshift from $z>5.7$ to $z<5.7$. An enhancement in IGM transmission than the average IGM transmission is observed at 5~cMpc from $5.7<z<6.14$ \oiii\ emitters, while around $5.3<z<5.7$ \oiii\ emitters, the IGM transmission is monotonically increasing with distance, up to the average IGM transmission; And (2) as shown in Figure \ref{fig:IGM_tau_eff_o3e}, $\tau_{\rm eff,[OIII]}$ measurements at $z>5.7$ have nearly the same integrated length when measuring the effective optical depth around \oiii\ emitters. The nearly uniform integrated length will ensure a consistent scale when we compare the IGM transmission around \oiii\ emitters with the control sample.

\subsubsection{CDFs at $5.7<z<6.1$}

Figure \ref{fig:cdf_different scales_high_z} shows the ``optimistic" and ``pessimistic" CDFs of $\tau_{\rm eff,[OIII]}$ around $5.7<z<6.1$ \oiii\ emitters. We also use the KM estimator to fit the survival function of each distribution. The fitted CDF and the corresponding $1\sigma$ confidence interval is shown as the shaded region in Figure \ref{fig:cdf_different scales_high_z}. Similar to the trend observed among all $z<6.1$ $\tau_{\rm eff,[OIII]}$ measurements, the fitted CDF of $\tau_{\rm eff,[OIII]}$ is higher than the fitted random CDF at the same $\tau_{\rm eff,[OIII]}$, suggesting the $\tau_{\rm eff}$ around \oiii\ emitters is generally lower. The most noticeable difference between the fitted CDFs appears at the right tail ($\tau_{\rm eff,[OIII]}>4$) of the distribution. The significant difference appears at a smaller scale (with an influence radius $\gtrsim20~h^{-1}\;\!$cMpc) with a p-value $<0.05$ from the log-rank test and is not smeared out at large scales (an influence radius $\sim50~h^{-1}\;\!$cMpc). This suggests that IGM patches are more transparent around \oiii\ emitters on scales $\gtrsim20~h^{-1}\;\!$cMpc, implying ionizing photon contribution from \oiii\ emitters to the local ionizing background. Similar excess IGM transmission around $z\sim6$ galaxies has been observed in \citet{Kakiichi2018MNRAS}, \citet{Meyer2019MNRAS,Meyer2020MNRAS}, and \citet{Kashino2023ApJ} by analyzing the IGM-galaxy cross-correlation function. \citet{Kakiichi2018MNRAS} find an excess IGM transmission at $\sim20$~cMpc from LBGs. \citet{Meyer2019MNRAS} find an excess IGM transmission at $>10~h^{-1}\;\!$cMpc for faint galaxies traced by \ion{C}{4} absorbers and \citet{Meyer2020MNRAS} report an excess IGM transmission at $>10$~cMpc around LBGs and LAEs. Using $5.7<z<6.14$ \oiii\ emitters, \citet{Kashino2023ApJ} find a peak of IGM transmission at $\sim5$~cMpc. As these studies use various galaxy populations when measuring the IGM-galaxy cross-correlation function, the scales where the excess IGM transmission can be different due to different galaxy populations residing in different IGM environments \citep{Momose2021ApJ}. In addition, \citet{Kakiichi2018MNRAS} and \citet{Kashino2023ApJ} measures the IGM-galaxy cross-correlation function in a single quasar field, therefore, the results can be subject to cosmic variance. 

\begin{comment}
Noticeably, at small scales $\sim5~h^{-1}\;\!$cMpc, $\tau_{\rm eff,[OIII]}$ shows more lower limits in the CDF than the random distribution shown in the brown line, potentially indicating that the IGM transmission can be suppressed around $5.7<z<6.1$ \oiii\ emitters at small scales $\sim5~h^{-1}\;\!$cMpc. However, due to the depth of the current data, the CDFs do not distinguish the difference between $\tau_{\rm eff,[OIII]}$ and the random distribution. Significantly deeper Ly$\alpha$ forest spectra from ground-based 6$-$10\;\!m telescopes will be beneficial to investigate the IGM transmission around $z>5.7$ \oiii\ emitters within small scales of $\sim5~h^{-1}\;\!$cMpc.
\end{comment}

\subsubsection{CDFs at $5.4<z<5.7$}

Figure \ref{fig:cdf_different scales_low_z} shows the CDFs of $\tau_{\rm eff,[OIII]}$ around $5.4<z<5.7$ \oiii\ emitters. With an influence radius of $5~h^{-1}\;\!$cMpc, the ``optimistic" CDF of $\tau_{\rm eff,[OIII]}$ shows a lower cumulative probability at the same $\tau_{\rm eff,[OIII]}$ than the random distribution, indicating the IGM is more opaque close to $z<5.7$ \oiii\ emitters within a scale of $5~h^{-1}\;\!$cMpc. However, from the KM estimator, the $1\sigma$ confidence intervals of two fitted CDFs overlap, and the p-value from the log-rank test is $0.33$, indicating Ly$\alpha$ absorption near \oiii\ emitters is not significant, based on the current CDFs of $\tau_{\rm eff,[OIII]}$. This can be caused by the large uncertainties in the fitted CDF, likely associated with the limited number of $z<5.7$ $\tau_{\rm eff,[OIII]}$ measurements. Ly$\alpha$ absorption near galaxies has been observed in previous studies. By cross-correlating \ion{C}{4} absorbers at $4.5<z<6.2$ and Ly$\alpha$ forest, \citet{Meyer2019MNRAS} find Ly$\alpha$ absorption within $5~h^{-1}\;\!$cMpc from \ion{C}{4} absorbers, indicating highly opaque regions surrounding faint galaxies traced by \ion{C}{4} absorbers. \citet{Kashino2023ApJ} measure the IGM-galaxy cross-correlation function through \oiii\ emitters and Ly$\alpha$ forest, and find IGM Ly$\alpha$ absorption within 8~cMpc from $5.3<z<5.7$ \oiii\ emitters. \citet{Tang2024arXiv} find that the Ly$\alpha$ profile of $z\sim5-6$ strongest LAEs displays a typical redshifted Ly$\alpha$ velocity of 230~${\rm km\;\!s^{-1}}$ compared with H$\alpha$ emission line, and the Ly$\alpha$ profile does not show the blue Ly$\alpha$ peak as observed among $z\sim2-3$ LAEs, suggesting the existence of residual \ion{H}{1} around galaxies in the IGM absorbs the blue component of Ly$\alpha$ emission. At $z\sim2-3$, similar Ly$\alpha$ forest absorption has also been found around several galaxy populations potentially due to \ion{H}{1} overdensity around galaxies \cite[e.\,g.,\,][]{Momose2021ApJ} or galaxy clusters \cite[e.\,g.,\,][]{Cai2017ApJ}. At $z\sim2-3$, \citet{Mukae2017ApJ} find a weak anti-correlation between the IGM transmission and galaxy overdensity within $2.5$ proper Mpc. From the current CDFs of $\tau_{\rm eff,[OIII]}$, Ly$\alpha$ absorption within $5~h^{-1}\;\!$cMpc from $z\sim5.4-5.7$ \oiii\ emitters is not significant. More $z<5.7$ $\tau_{\rm eff,[OIII]}$ measurements will be needed to reconstruct the $\tau_{\rm eff,[OIII]}$ distribution and to verify the existence of Ly$\alpha$ absorption near \oiii\ emitters within $5~h^{-1}\;\!$cMpc. 
%\textcolor{blue}{\textbf{Lower $\tau_{\rm eff,[OIII]}$ within $5~h^{-1}\;\!$cMpc around $z\sim5.4-5.7$ \oiii\ emitters suggests Ly$\alpha$ absorption due to \ion{H}{1} gas overdensity around \oiii\ emitters, and the scale showing absorption is consistent with existing studies at $z\sim5-6$ \citep{Meyer2019MNRAS,Kashino2023ApJ}.}}

When adopting a larger influence radius than $5~h^{-1}\;\!$cMpc, the fitted CDF of $\tau_{\rm eff,[OIII]}$ from the KM estimator first tends to be consistent with the fitted random CDF. However, with an influence radius $\gtrsim25~h^{-1}\;\!$cMpc, the fitted CDF of $\tau_{\rm eff,[OIII]}$ is higher than the fitted random CDF at the same $\tau_{\rm eff,[OIII]}$, displaying a significant enhancement of IGM transmission around \oiii\ emitters on a larger scale $\gtrsim25~h^{-1}\;\!$cMpc. The fitted CDF of $\tau_{\rm eff,[OIII]}$ shows substantial divergence from the fitted random CDF at $\tau_{\rm eff,[OIII]}\sim3$. 
%Because the number of quasars covering $z<5.7$ Ly$\alpha$ forest is limited in this study, and $\tau_{\rm eff,[OIII]}$ can have limited integrated length because some \oiii\ emitters are at the boundary of the Ly$\alpha$ forest, we emphasize that the evidence of excess IGM transmission around $5.4<z<5.7$ \oiii\ emitters on scales $\gtrsim25~h^{-1}\;\!$cMpc is tentative. 
We further search the evidence of excess IGM transmission at $z<5.7$ using existing WFSS programs in quasar fields which fully cover $5.4<z<5.7$ Ly$\alpha$ forest. In \citet{Kashino2023ApJ}, the IGM-galaxy cross-correlation function beyond $15$~cMpc is not available. By utilizing the published \oiii\ emitter catalog in \citet{Kashino2023ApJ} and the optical spectrum of SDSS J0100$+$2802 from XQR-30 Github repository, we compute the IGM-galaxy cross-correlation function of $5.3<z<5.7$ \oiii\ emitters to a larger scale and find at $\sim20$~cMpc from $5.3<z<5.7$ \oiii\ emitters, there is also an enhanced IGM transmission compared with the average IGM transmission. However, since there is only one single quasar field, the result is greatly affected by cosmic variance. 
%Those results highlight the need for more galaxy redshift surveys in quasar fields, especially at $z<6$, to further verify the existence of enhanced IGM transmission around $z<5.7$ galaxies on large scales.  

For both $5.7<z<6.1$ and $5.4<z<5.7$ bins, IGM transmission is observed to be higher around \oiii\ emitters on large scales. The significant enhanced IGM transmission starts to appear at $\gtrsim20~h^{-1}\;\!$cMpc around $z>5.7$ \oiii\ emitters than $\gtrsim25~h^{-1}\;\!$cMpc scales of $z<5.7$ \oiii\ emitters. The scales where the excess IGM transmission emerges are similar for both $z>5.7$ and $z<5.7$ \oiii\ emitters, however, the distinction between the fitted CDF of $\tau_{\rm eff,[OIII]}$ and the fitted random CDF appears at $\tau_{\rm eff,[OIII]}>4$ for $z>5.7$ \oiii\ emitters, while for $z<5.7$ \oiii\ emitters, the fitted CDF of $\tau_{\rm eff,[OIII]}$ already deviates from the fitted random CDF in the most transparent regime ($\tau_{\rm eff,[OIII]}\lesssim3$). This is related with the rapid redshift evolution of $\tau_{\rm eff}$ and a significant portion of $z>5.7$ $\tau_{\rm eff,[OIII]}$ are lower limits, resulting in large uncertainty in the fitted CDF. To better probe the excess IGM transmission around \oiii\ emitters at $z>5.7$ and to further investigate whether the excess IGM transmission around \oiii\ emitters evolves with redshift, deep optical spectroscopy is needed (see also discussion in Section \ref{sec:trans_flux}). 

%both beyond the recent ionizing photon mean free path measurements at the corresponding redshift ($\lambda_{\rm mfp}=0.81~{\rm pMpc}$ ($5.61~{\rm cMpc}$) at $z=5.93$ and $\lambda_{\rm mfp}=3.31~{\rm pMpc}$ ($22.01~{\rm cMpc}$) at $z=5.65$, see \citealt{Zhu2023ApJ}).}} 
%The change in the scales where the excess IGM transmission emerges is similar to the rapid evolution of ionizing photon mean free path from $z\sim6$ to $z\sim5$ \citep{Becker2018ApJ,Zhu2023ApJ,Gaikwad2023MNRAS,Davies2023arXiv}. It is worth noting that the smallest scales where there is an significant enhancement of the IGM transmission around \oiii\ emitters are only slightly larger than the ionizing photon mean free path ($\lambda_{\rm mfp}$) at the corresponding redshift: $\lambda_{\rm mfp}=0.81~{\rm pMpc}$ ($\sim5.6~{\rm cMpc}$) at $z=5.93$ to $\lambda_{\rm mfp}=3.31~{\rm pMpc}$ ($\sim22.0~{\rm cMpc}$) at $z=5.65$ \citep{Zhu2023ApJ}, suggesting the large scatter in the observed $\tau_{\rm eff}$ is tightly associated with the fluctuations in the ionizing background. 

It is worth noting that the scales around \oiii\ emitters where excess IGM transmission emerges should not be interpreted as the size of ionized bubbles directly \cite[e.\,g.,\,][]{Tilvi2020ApJ,Endsley2022MNRAS,Tang2023MNRAS,Umeda2023arXiv,Napolitano2024arXiv,Whitler2024MNRAS,Neyer2024MNRAS}. Because $z\sim6$ is close to the end of reionization, the majority of ionized regions in the IGM have largely overlapped in lieu of individually expanding in the significantly neutral IGM in the early stage of reionization. Beyond the scales around \oiii\ emitters where excess IGM transmission exhibits, the IGM is already ionized, not significantly neutral. Excess IGM transmission around \oiii\ emitters indicates the enhancement in the ionizing photons close to \oiii\ emitters within certain scales, compared with the average ionizing background \cite[see also discussions in ][]{Fan2006AJ}.  

Because the FoV of JWST NIRCam is small, it is plausible that our identification of IGM patches where no \oiii\ emitters are detected is not complete. Further \oiii\ emitters located at a transverse distance of $\gtrsim6~h^{-1}\;\!$cMpc will not be covered by the existing single NIRCam pointing. 
%However, as demonstrated in Figure \ref{fig:cdf_different scales_high_z} and Figure \ref{fig:cdf_different scales_low_z}, the minimal scale showing significantly excess IGM transmission is $\sim10~h^{-1}\;\!$cMpc at $z>5.7$ versus $\sim25~h^{-1}\;\!$cMpc at $z<5.7$, therefore, the identification of IGM patches where no \oiii\ emitters are detected is expected to be more complete at $z>5.7$ than $z<5.7$. 
In addition, the current depth of ASPIRE program can only detect \oiii\ emitters down to a luminosity at $\sim 10^{42}~{\rm erg\;\!s^{-1}}$. It is also likely that IGM patches where no \oiii\ emitters are detected can still have fainter, undetected \oiii\ emitters around them within the FoV of a single NIRCam pointing. A follow-up JWST program (PID: 3325, PI: F. Wang and J. Yang) targets two ASPIRE quasars with a bigger mosaic ($\sim4.4'\times7.3'$, corresponding to $7.4~h^{-1}\;\!$cMpc$\times12.2~h^{-1}\;\!$cMpc at $z\sim6$) and a deeper exposure around the quasar with NIRCam WFSS. We will use \oiii\ emitters selected from ASPIRE follow-up programs to test the influence of both FoV and depth on our results. 

\subsection{Average transmitted flux around \oiii\ emitters}\label{sec:trans_flux}

\begin{figure*}
    \centering
    \includegraphics[width=1.0\textwidth]{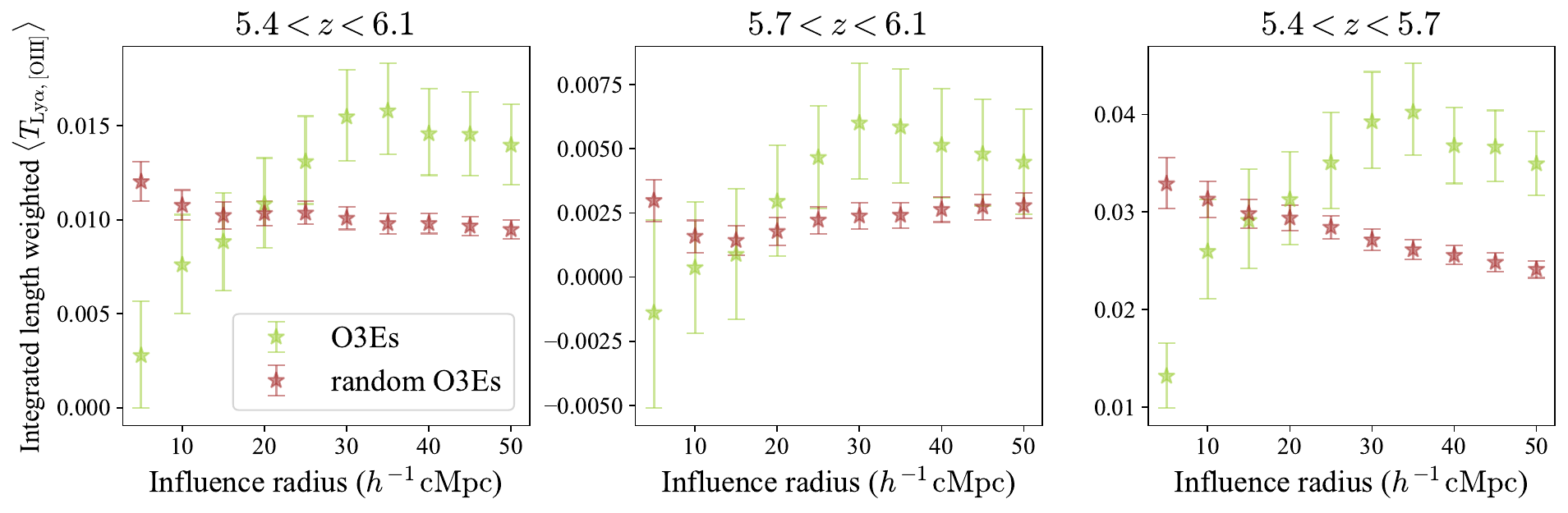}
    %\caption{The cumulative distribution functions of the IGM effective optical depth around \oiii\ emitters at $5.4<z<5.7$ (yellowgreen) and the random distribution (brown), adopting influence radii from 5~$h^{-1}\;\!{\rm cMpc}$ to 50~$h^{-1}\;\!{\rm cMpc}$. The $1\sigma$ uncertainty of the CDF derived from bootstrapping sampling is shown in the shaded regions of the corresponding color. The null-hypothesis p-value for AD tests is shown in the upper left corner of each subpanel. With an influence radius of 5~$h^{-1}\;\!{\rm cMpc}$, the IGM effective optical depth around \oiii\ emitters is significantly higher than the random distribution, suggesting Ly$\alpha$ absorption closed to \oiii\ emitters due to gas overdensity. Significant excess of IGM transmission around \oiii\ emitters is shown beyond an influence radius of 25~$h^{-1}\;\!{\rm cMpc}$.}
    \caption{Average transmitted flux around \oiii\ emitters (yellow green stars), weighted by the integrated length of each $\tau_{\rm eff,[OIII]}$ measurement, as a function of influence radii at $5.4<z<6.1$, $5.7<z<6.1$, and $5.4<z<5.7$. The $1\sigma$ uncertainty is calculated using bootstrapping. The random values are denoted by brown stars.}
    \label{fig:o3_transflux}
\end{figure*}

In each distribution of $\tau_{\rm eff,[OIII]}$, 
%due to the data quality of quasar optical spectroscopy, 
a subset of measurements are $2\sigma$ lower limits. In Section \ref{sec:discussion_CDF_blz6p1} and Section \ref{sec:discussion_CDF_z_evolve}, we adopt the KM estimator to fit the CDF of $\tau_{\rm eff,[OIII]}$ and perform the log-rank test to investigate whether the IGM patches around \oiii\ emitters tend to be more transparent or more opaque than average IGM transmission. To verify whether the results are subject to the fact that $2\sigma$ lower limits of $\tau_{\rm eff,[OIII]}$ are included in the distribution, we calculate the average transmitted flux in the distribution, and explore whether there is excess IGM transmission compared with the random distribution, which represents a baseline of IGM transmission. 
%We evaluate the impact of including $2\sigma$ lower limits as measurements in the distribution by calculating the average transmitted flux in the distribution. 
For each distribution in Figure \ref{fig:cdf_different scales}, \ref{fig:cdf_different scales_high_z}, and \ref{fig:cdf_different scales_low_z}, we calculate its average transmitted flux when adopting a certain influence radius, weighted by the integrated length for each measurement, and derive the $1\sigma$ uncertainty in the average transmitted flux by bootstrapping. 
%10000 times. 
Figure \ref{fig:o3_transflux} shows the average transmitted flux around \oiii\ emitters as a function of influence radii in 3 redshift ranges: $5.4<z<6.1$, $5.7<z<6.1$, and $5.4<z<5.7$. It is important to note that Figure \ref{fig:o3_transflux} is different than the average transmitted flux as a function of distance to galaxies (i.\,e.,\,$T(r)$, see \citealt{Kakiichi2018MNRAS,Meyer2019MNRAS,Meyer2020MNRAS,Kashino2023ApJ}), because we calculate the average transmitted flux over the integrated length by adopting an influence radius, instead of at a certian distance from galaxies. 
%(please note this figure is different than the IGM-galaxy CCF because the flux was calculated within a radius instead of at a radius).

At $5.4<z<6.1$, the transmitted flux around \oiii\ emitters tends to be lower than the random distribution with an influence radius of $5~h^{-1}\;\!$cMpc. Compared with the random distribution, the excess in transmitted flux around \oiii\ emitters starts to appear with influence radii $\gtrsim25~h^{-1}\;\!$cMpc, and the excess in transmitted flux becomes more significant ($\sim2\sigma$) adopting an influence radius $\gtrsim30~h^{-1}\;\!$cMpc. The $25~h^{-1}\;\!$cMpc influence radius where the excess IGM transmission appears is consistent with the influence radius where the fitted CDF of $\tau_{\rm eff,[OIII]}$ is significantly different than the random distribution shown in the Figure \ref{fig:cdf_different scales}. 

As for the $5.7<z<6.1$ bin, with an influence radius of $\lesssim20~h^{-1}\;\!$cMpc, the average transmitted flux around \oiii\ emitters and the random transmitted flux overlap with each other within 1$\sigma$ uncertainty. Excess in transmitted flux around \oiii\ emitters exists when adopting influence radii $\gtrsim25~h^{-1}\;\!$cMpc. Such excess is most evident when adopting an influence radius $\sim30-35~h^{-1}\;\!$cMpc, but the significance of the excess in transmitted flux is less than $2\sigma$. Because the transmitted flux in the Ly$\alpha$ forest is low at 
$z\sim6$, it is challenging to constrain the transmitted flux precisely, especially using shallow spectroscopy. Therefore, the significance of excess transmission can be underestimated, because noisy spectra can dilute the signal of excess transmission.

For the $5.4<z<5.7$ bin, with an influence radius of $5~h^{-1}\;\!$cMpc, the average transmitted flux around \oiii\ emitters is significantly lower ($\sim4.7\sigma$) than the random value. However, the fitted CDF of $\tau_{\rm eff,[OIII]}$ from the KM estimator does not differ significantly from the random CDF when adopting an influence radius of $5~h^{-1}\;\!$cMpc, because of the large uncertainty in the fitted CDF. When increasing the influence radii to $\gtrsim25~h^{-1}\;\!$cMpc, the average transmitted flux around \oiii\ emitters become higher than the random average transmitted flux. The excess in transmitted flux around \oiii\ emitters is most significant ($\sim3.1\sigma$) when adopting influence radii $\sim30-50~h^{-1}\;\!$cMpc. The scales showing Ly$\alpha$ excess Ly$\alpha$ transmission are consistent with the scales where the fitted CDF of $\tau_{\rm eff,[OIII]}$ is significantly different than the random distribution in Figure \ref{fig:cdf_different scales_low_z}. 

To summarize, we find that the scales exhibiting lower or higher transmitted flux around \oiii\ emitters are mostly consistent with those where the fitted $\tau_{\rm eff,[OIII]}$ distribution from the KM estimator deviates significantly from the fitted random distribution. When investigating the transmission around \oiii\ emitters, both the average transmitted flux and the $\tau_{\rm eff,[OIII]}$ distribution exhibit their own advantages and limitations. Using the average transmitted flux around \oiii\ emitters, it is straightforward to determine whether the transmission is enhanced or suppressed around \oiii\ emitters, compared with the average IGM transmission. 
On the other hand, for $\tau_{\rm eff,[OIII]}$ measurements, a substantial sample size is required to reconstruct the distribution and thus to investigate transmission around \oiii\ emitters.
However, average transmitted flux can be biased towards high transmitted flux. Because the average IGM transmitted flux decreases substantially as the redshift increases \citep{Becker2015MNRAS,Bosman2018MNRAS,Bosman2022MNRAS,Eilers2018ApJ,Yang2020ApJb}, the average transmitted flux within a broad redshift range will be naturally biased towards lower redshift, while $\tau_{\rm eff,[OIII]}$ measurements can demonstrate the entire distribution of IGM transmission. Nevertheless, both methods can be limited by noisy spectra. Using noisy data, it is difficult to measure the transmitted flux precisely, and there are also more lower limits in the $\tau_{\rm eff,[OIII]}$ distribution, resulting in a large uncertainty in the fitted CDF.

Recent ionizing photon mean free path ($\lambda_{\rm mfp}$) measurements show the rapid evolution of $\lambda_{\rm mfp}=0.81^{+0.73}_{-0.48}~{\rm pMpc}$ at $z=5.93$ to $\lambda_{\rm mfp}=3.31^{+2.74}_ {-1.34}~{\rm pMpc}$ at $z=5.65$ (\citealt{Zhu2023ApJ}, see also \citealt{Becker2018ApJ,Gaikwad2023MNRAS,Davies2024ApJ}). In the scenario where the scatter in $\tau_{\rm eff}$ is primarily driven by ionizing background fluctuations, the fluctuations in the Ly$\alpha$ forest are stronger with a shorter $\lambda_{\rm mfp}$ \citep{Davies2024ApJ}. As such, the excess transmission around \oiii\ emitters might show a redshift evolution when the the scatter in $\tau_{\rm eff}$ is dominated by the ionizing background fluctuations. However, due to the moderate depth of data, currently we do not find conclusive evidence of the redshift evolution of excess transmission around \oiii\ emitters. 

%except for the $5.7<z<6.1$ redshift range: the CDF of $\tau_{\rm eff,[OIII]}$ is already significantly different than the random distribution adopting an influence radius of $10~h^{-1}\;\!$cMpc, while the average transmitted flux around \oiii\ emitters displays excess than the random value only with influence radii $\gtrsim25~h^{-1}\;\!$cMpc.
%Because the transmitted flux is low at $z\sim6$, it is challenging to constrain the transmitted flux precisely, and a significant portion of $\tau_{\rm eff,[OIII]}$ are also in the format of $2\sigma$ lower limits. \textcolor{red}{revise the discussion here} By treating the $2\sigma$ lower limits of $\tau_{\rm eff,[OIII]}$ as measurements in the distribution, we would get an ``optimistic" estimation of the transmitted flux around \oiii\ emitters. However, by calculating the average transmitted flux around \oiii\ emitters, we could also underestimate the significance of excess transmission, because noisy spectra can dilute the signal of excess transmission. 
This emphasizes the necessity of deeper exposure and more quasar sightlines for quasar optical spectroscopy, especially to detect weak transmission at $z\sim6$, in order to investigate the redshift evolution of the excess IGM transmission around \oiii\ emitters. 
%This does not imply that excess transmission does not exist within $10~h^{-1}\;\!$cMpc from $5.7<z<6.1$ \oiii\ emitters. As the transmitted flux is very low (\textcolor{red}{add number}) at $z\sim6$, it is difficult to measure the transmitted flux well and resolve weak transmission spikes in relatively shallow quasar Ly$\alpha$ forest spectra. 

% Similar excess in transmitted flux around \oiii\ emitters can also be found in both $5.7<z<6.1$ and $5.4<z<5.7$ bins with influence radii $\gtrsim25~h^{-1}\;\!$cMpc. 
% For the $5.7<z<6.1$ bin, the scale where the excess transmitted flux emerges is larger than the scale derived from the CDF. While for the $5.4<z<5.7$ bin, the excess Ly$\alpha$ transmission appear on the scales. 

% Tentative evidence of lower transmitted flux within $\gtrsim5~h^{-1}\;\!$cMpc from $5.7<z<6.1$ \oiii\ emitters, while for $5.4<z<5.7$, the absorption within 5cMpc/h is pretty significant. For the $5.4<z<5.7$ bin, the scale where Ly$\alpha$ absorption and excess Ly$\alpha$ transmission appears is consistent with the scales derived from CDF. However, for the $5.7<z<6.1$ bin, using an influence radius of 10 cMpc/h, transmission flux around \oiii\ emitters is not significantly higher than the random distribution. Only at scales $\gtrsim$25cMpc/h, the excess transmission emerges, but the significance is less than $2\sigma$ based on the current data quality. This highlights the need for deeper spectroscopy for ASPIRE quasars, especially to detect weak transmission, in order to detect the excess transmission. 

%{\color{red}{Discussion about the mean free path? $\lambda_{\rm mfp} (z=5.65) =3.31~{\rm pMpc}$ and $\lambda_{\rm mfp} (z=5.93) =~0.81 {\rm pMpc}$}}

\section{Summary} \label{sec:conclusion}
%% IMPORTANT! The old "\acknowledgment" command has be depreciated. It was
%% not robust enough to handle our new dual anonymous review requirements and
%% thus been replaced with the acknowledgment environment. If you try to 
%% compile with \acknowledgment you will get an error print to the screen
%% and in the compiled pdf.
%% 
%% Also note that the akcnowlodgment environment does not support long amounts of text. If you have a lot of people and institutions to acknowledge, do not use this command. Instead, create a new \section{Acknowledgments}.
In this paper, we present the first $\tau_{\rm eff}$ measurements around \oiii\ emitters at $5.4<z<6.1$ identified in 14 fields of the ASPIRE program centered on $z\sim6.5-6.8$ quasars for which we have a good quality spectrum. We find: 
%using 14 ASPIRE quasar sightlines and $5.4<z<6.1$ \oiii\ emitters identified in quasar fields using ASPIRE program. We find:
\begin{itemize}
    \item By stacking the Ly$\alpha$ transmission around \oiii\ emitters with an influence radius of $25~h^{-1}\;\!$cMpc, we find at the same redshift, IGM patches around \oiii\ emitters are significantly more transparent (${\rm d}\tau_{\rm eff}>0.5$) than those IGM patches where no \oiii\ emitters are detected. At $z>5.7$, most IGM transmission is within $25~h^{-1}\;\!$cMpc from \oiii\ emitters, reinforcing the important roles of galaxies in producing IGM transmission as observed in the Ly$\alpha$ forest. 
    \item Stacked IGM patches around \oiii\ emitters reach the same optical depth at least ${\rm d}z\sim0.1$ ahead of stacked IGM patches away from \oiii\ emitters, supporting earlier reionization around \oiii\ emitters. 
    \item 
    %By splitting all \oiii\ emitters into two redshift bins: $5.4<z<5.7$ and $5.7<z<6.1$, 
    With an influence radius $\gtrsim25~h^{-1}\;\!$cMpc, we find that the $\tau_{\rm eff}$ distribution around $5.4<z<6.1$ \oiii\ emitters is significantly different from the IGM $\tau_{\rm eff}$ distribution, and the average transmitted flux around \oiii\ emitters is higher than the average IGM transmitted flux, supporting that the scatter in the observed $z>5.5$ $\tau_{\rm eff}$ is tightly associated with the large-scale fluctuations in the ionizing background. By splitting all \oiii\ emitters into two redshift bins: $5.4<z<5.7$ and $5.7<z<6.1$, we find the scales where the excess IGM transmission emerges are similar for $5.4<z<5.7$ ($\gtrsim25~h^{-1}\;\!$cMpc) and $5.7<z<6.1$ ($\gtrsim20~h^{-1}\;\!$cMpc) bins. Deeper optical spectroscopy will be needed to further investigate the redshift evolution of excess transmission around \oiii\ emitters. 
\end{itemize} 

%By splitting all \oiii\ emitters into two redshift bins: $5.4<z<5.7$ and $5.7<z<6.1$, we find the IGM effective optical depth measurements around \oiii\ emitters are significantly lower than the average IGM effective optical depth distribution beyond the ionizing photon mean free path at corresponding redshifts, supporting that the significant scatter in the observed $z>5.5$ IGM effective optical depth is mainly driven by the large-scale fluctuations in the ionizing background. This also suggests that \oiii\ emitters can contribute significantly to the local ionizing background. 

We have ongoing programs on 6-10m ground-based telescopes to collect the optical spectroscopy of ASPIRE quasars to complete the ASPIRE sample and also to increase the S/N of the existing optical quasar spectroscopy. The IGM-galaxy connection studies of the full ASPIRE quasar sample will be presented in later works.

\section*{Acknowledgements}

%{\color{blue}{funding:}} 
We thank the anonymous reviewer for constructive comments that improved this manuscript. XJ thanks Yongda Zhu, Daichi Kashino, and Andrei Mesinger for informative discussions. 
FW acknowledges support from NSF Grant AST-2308258.
RAM acknowledges support from the Swiss National Science Foundation (SNSF) through project grant 200020\_207349. GDB was supported by the NSF through grant AST-1751404. SZ acknowledges support from the National Science Foundation of China (grant no.~12303011). 

%{\color{blue}{data:}} 
This work is based on observations made with the NASA/ESA/CSA James Webb Space Telescope. The data were obtained from the Mikulski Archive for Space Telescopes at the Space Telescope Science Institute, which is operated by the Association of Universities for Research in Astronomy, Inc., under NASA contract NAS 5-03127 for JWST. The specific observations analyzed can be accessed via \dataset[10.17909/vt74-kd84]{https://doi.org/10.17909/vt74-kd84}. These observations are associated with program \#2078. Support for program \#2078 was provided by NASA through a grant from the Space Telescope Science Institute, which is operated by the Association of Universities for Research in Astronomy, Inc., under NASA contract NAS 5-03127. 

%VLT
This work is based in part on observations made with ESO telescopes at the La Silla Paranal Observatory under program IDs 
% J0109-3047
087.A-0890(A), 088.A-0897(A), 
%J1048m0109
097.B-1070(A), 
%J0305 
098.A-0444(A), 
%J2232
098.B-0537(A),
%J0224 & J0226 & J0923+0402 & J
0100.A-0625(A), 0102.A-0154(A), 1103.A-0817(A), 1103.A-0817(B), 
%J1129
2102.A-5042(A).
%084.A-0360(A), 084.A-0390(A), 084.A-0550(A), 085.A-0299(A), 086.A-0162(A), 087.A-0607(A), 096.A-0095(A), 096.A-0418(A), 097.B-1070(A), 098.A-0444 (A), 098.A-0527(A), 098.B-0537(A), 0100.A-0446(A), 0100.A-0625(A), 0102.A-0154(A), and 0103.A-0423(A). 
% Gemini
The paper also used data based on observations obtained at the international Gemini Observatory, a program of NSF's NOIRLab, which is managed by the Association of Universities for Research in Astronomy (AURA) under a cooperative agreement with the National Science Foundation. on behalf of the Gemini Observatory partnership: the National Science Foundation (United States), National Research Council (Canada), Agencia Nacional de Investigaci\'{o}n y Desarrollo (Chile), Ministerio de Ciencia, Tecnolog\'{i}a e Innovaci\'{o}n (Argentina), Minist\'{e}rio da Ci\^{e}ncia, Tecnologia, Inova\c{c}\~{o}es e Comunica\c{c}\~{o}es (Brazil), and Korea Astronomy and Space Science Institute (Republic of Korea). 
% Magellan
This paper includes data gathered with the 6.5 meter Magellan Telescopes located at Las Campanas Observatory, Chile.
% Keck 
Some of the data presented herein were obtained at the W. M. Keck Observatory, which is operated as a scientific partnership among the California Institute of Technology, the University of California and the National Aeronautics and Space Administration. The Observatory was made possible by the generous financial support of the W. M. Keck Foundation. 

%{\color{blue}{land:}} 
The authors wish to recognize and acknowledge the very significant cultural role and reverence that the summit of Maunakea has always had within the indigenous Hawaiian community. We are most fortunate to have the opportunity to conduct observations from this mountain. 

We respectfully acknowledge the University of Arizona is on the land and territories of Indigenous peoples. Today, Arizona is home to 22 federally recognized tribes, with Tucson being home to the O'odham and the Yaqui. Committed to diversity and inclusion, the University strives to build sustainable relationships with sovereign Native Nations and Indigenous communities through education offerings, partnerships, and community service.

%% To help institutions obtain information on the effectiveness of their 
%% telescopes the AAS Journals has created a group of keywords for telescope 
%% facilities.
%
%% Following the acknowledgments section, use the following syntax and the
%% \facility{} or \facilities{} macros to list the keywords of facilities used 
%% in the research for the paper.  Each keyword is check against the master 
%% list during copy editing.  Individual instruments can be provided in 
%% parentheses, after the keyword, but they are not verified.

\vspace{5mm}
\facilities{JWST, VLT (X-Shooter), Gemini (GMOS), Keck (LRIS, DEIMOS), Magellan (LDSS3)}

%% Similar to \facility{}, there is the optional \software command to allow 
%% authors a place to specify which programs were used during the creation of 
%% the manuscript. Authors should list each code and include either a
%% citation or url to the code inside ()s when available.

\software{astropy \citep{2013A&A...558A..33A,2018AJ....156..123A,astropy:2022}, Numpy \citep{Numpy2020}, Matplotlib \citep{Matplotlib2007}, PypeIt \citep{Prochaska2020JOSS,Prochaska2020zndo}, Scipy \citep{2020SciPy-NMeth}
          }

%% Appendix material should be preceded with a single \appendix command.
%% There should be a \section command for each appendix. Mark appendix
%% subsections with the same markup you use in the main body of the paper.

%% Each Appendix (indicated with \section) will be lettered A, B, C, etc.
%% The equation counter will reset when it encounters the \appendix
%% command and will number appendix equations (A1), (A2), etc. The
%% Figure and Table counter will not reset.

\appendix

\section{Quasar Sightline Plots}\label{appendix:los}

In this section, we show the transmitted spectrum of all ASPIRE quasar sightlines used in this work and the spatial location of \oiii\ emitters identified in the quasar fields in Figure \ref{fig:los_plot_1}, Figure \ref{fig:los_plot_2}, Figure \ref{fig:los_plot_3}, and Figure \ref{fig:los_plot_4}. 

\begin{comment}
\begin{figure*}[!ht]
    \centering
    \includegraphics[width=0.95\textwidth]{figures/J0109m3047.pdf}
    \includegraphics[width=0.95\textwidth]{figures/J0218p0007.pdf}
    \includegraphics[width=0.95\textwidth]{figures/J0224m4711.pdf}
    \includegraphics[width=0.95\textwidth]{figures/J0226p0302.pdf}
    \includegraphics[width=0.95\textwidth]{figures/J0244m5008.pdf}
    \includegraphics[width=0.95\textwidth]{figures/J0305m3150.pdf}
    \includegraphics[width=0.95\textwidth]{figures/J0525m2406.pdf}
    \includegraphics[width=0.95\textwidth]{figures/J0706p2921.pdf}
    \includegraphics[width=0.95\textwidth]{figures/J0923p0402.pdf}
    \includegraphics[width=0.95\textwidth]{figures/J1048m0109.pdf}
    \includegraphics[width=0.95\textwidth]{figures/J1104p2134.pdf}
    \includegraphics[width=0.95\textwidth]{figures/J1129p1846.pdf}
    \includegraphics[width=0.95\textwidth]{figures/J1526m2050.pdf}
    \includegraphics[width=0.95\textwidth]{figures/J2002m3013.pdf}
    \includegraphics[width=0.95\textwidth]{figures/J2102m1458.pdf}
    \includegraphics[width=0.95\textwidth]{figures/J2232p2930.pdf}
    \caption{The line of sight plot of each quasar.}
    \label{fig:los_plot}
\end{figure*}
\end{comment}

\begin{figure*}[!ht]
    \centering
    \includegraphics[width=0.95\textwidth]{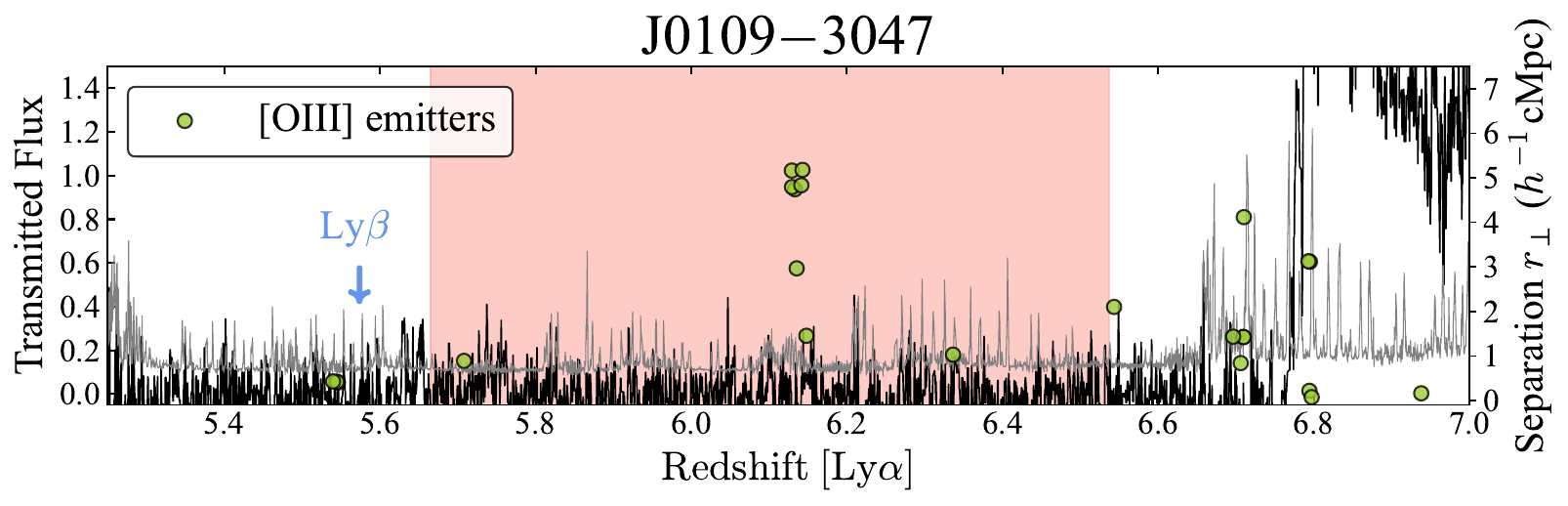}
    \includegraphics[width=0.95\textwidth]{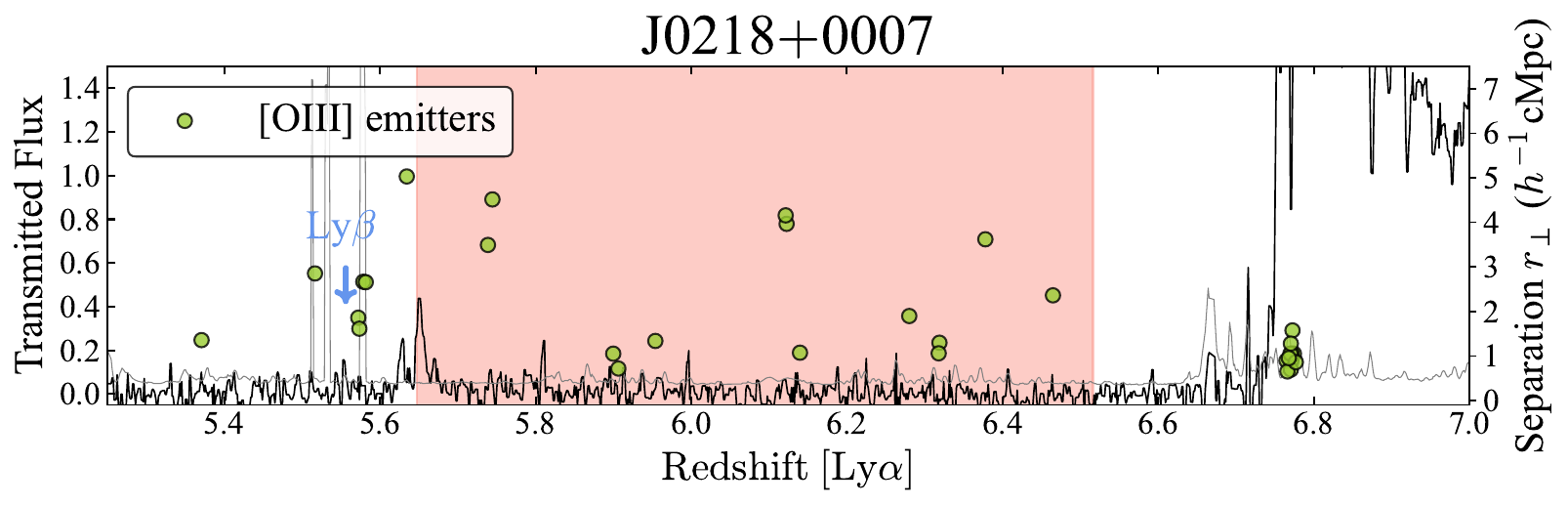}
    \includegraphics[width=0.95\textwidth]{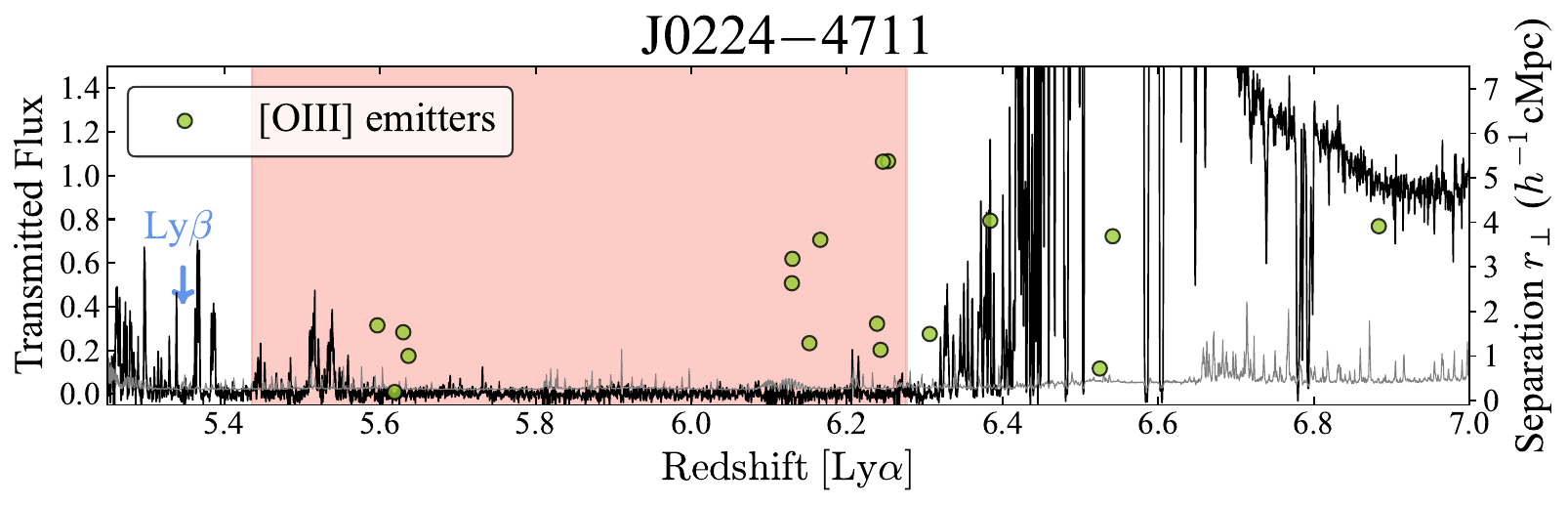}
    \includegraphics[width=0.95\textwidth]{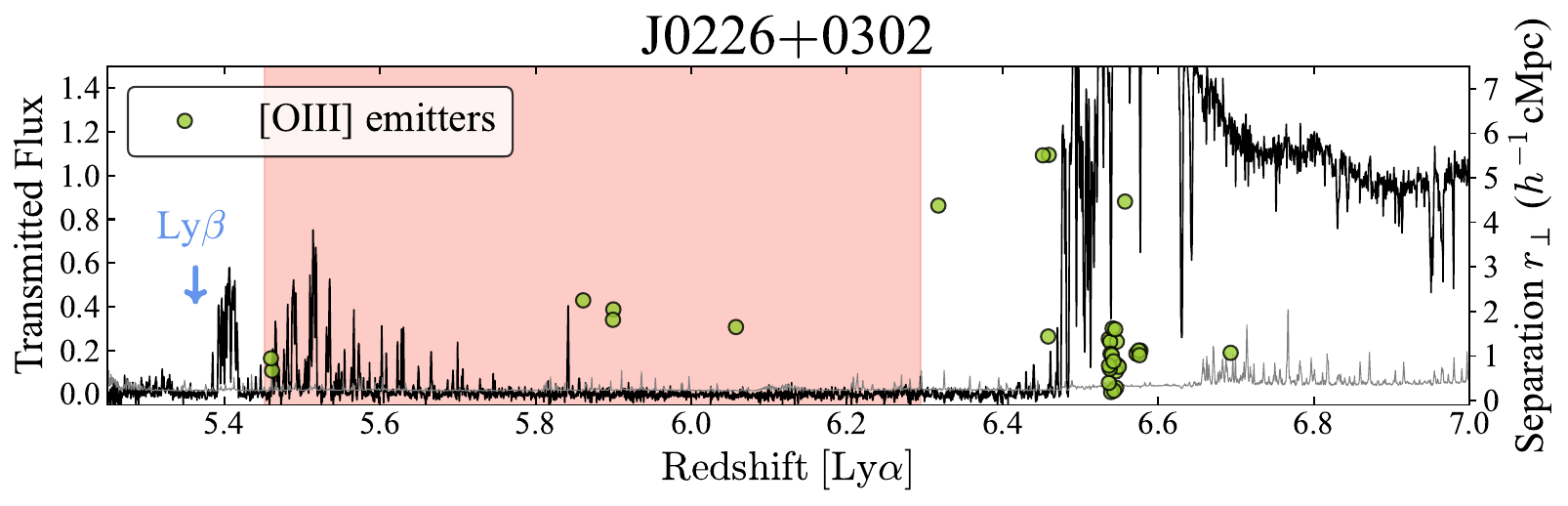}
    \caption{The transmitted spectrum of all quasars used in this work, in the ascending order of R.A. The transmitted spectrum (black) has been smoothed with a median filter for better visualization. The spectral uncertainty is shown in gray. The redshift range of the Ly$\alpha$ used in the analysis is displayed in the red shaded region. The location of Ly$\beta$ emission line is marked by the blue downwards arrow. The spatial locations of the \oiii\ emitters identified in each quasar field are denoted by the yellowgreen circles in terms of the \oiii\ emitter redshift $z_{\rm [OIII]}$ and the transverse distance $r_{\rm \perp}$ between the \oiii\ emitter and the central quasar. }
    \label{fig:los_plot_1}
\end{figure*}

\begin{figure*}[!ht]
    \centering
    \includegraphics[width=0.95\textwidth]{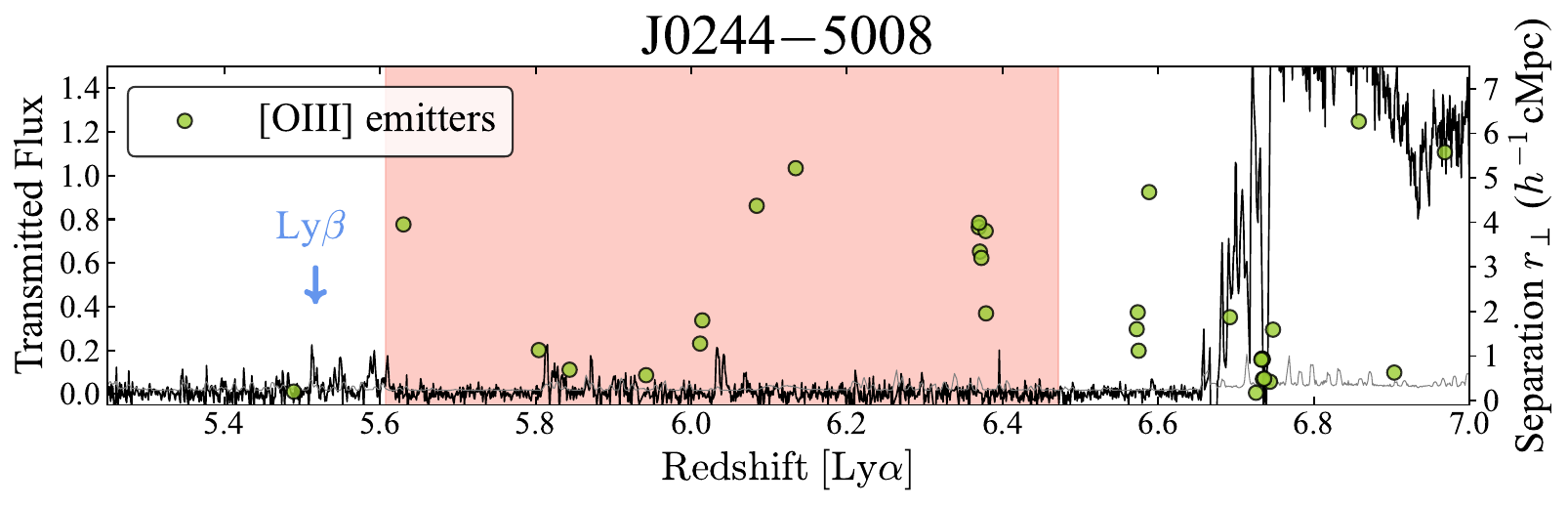}
    \includegraphics[width=0.95\textwidth]{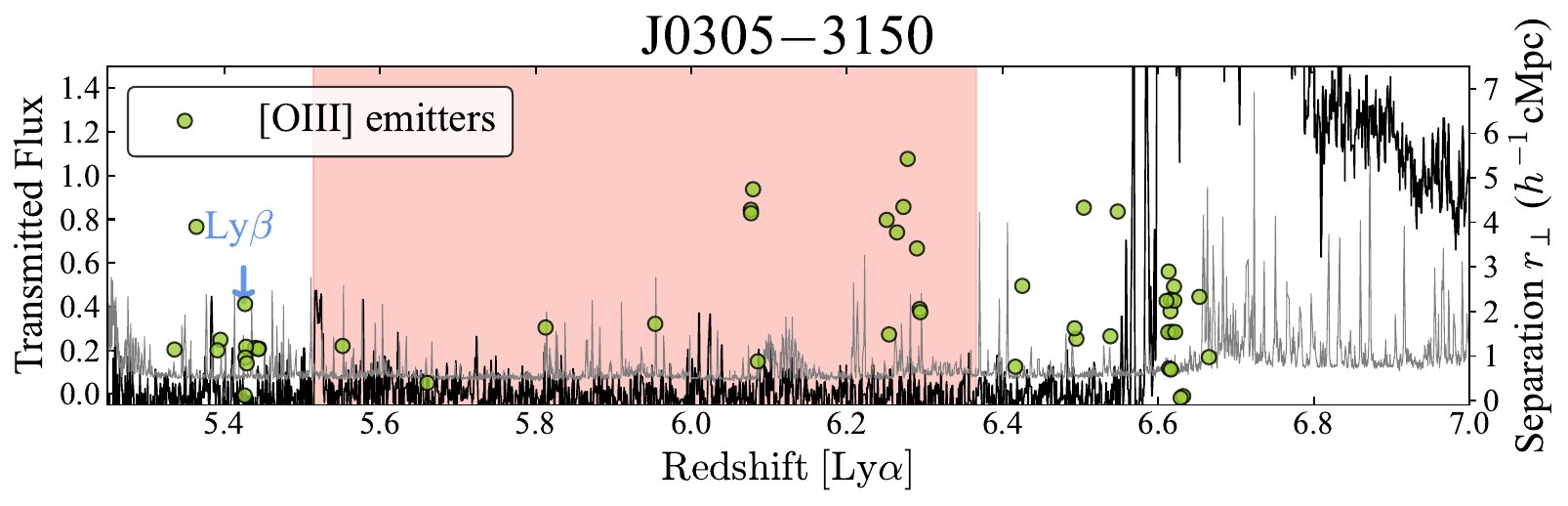}
    \includegraphics[width=0.95\textwidth]{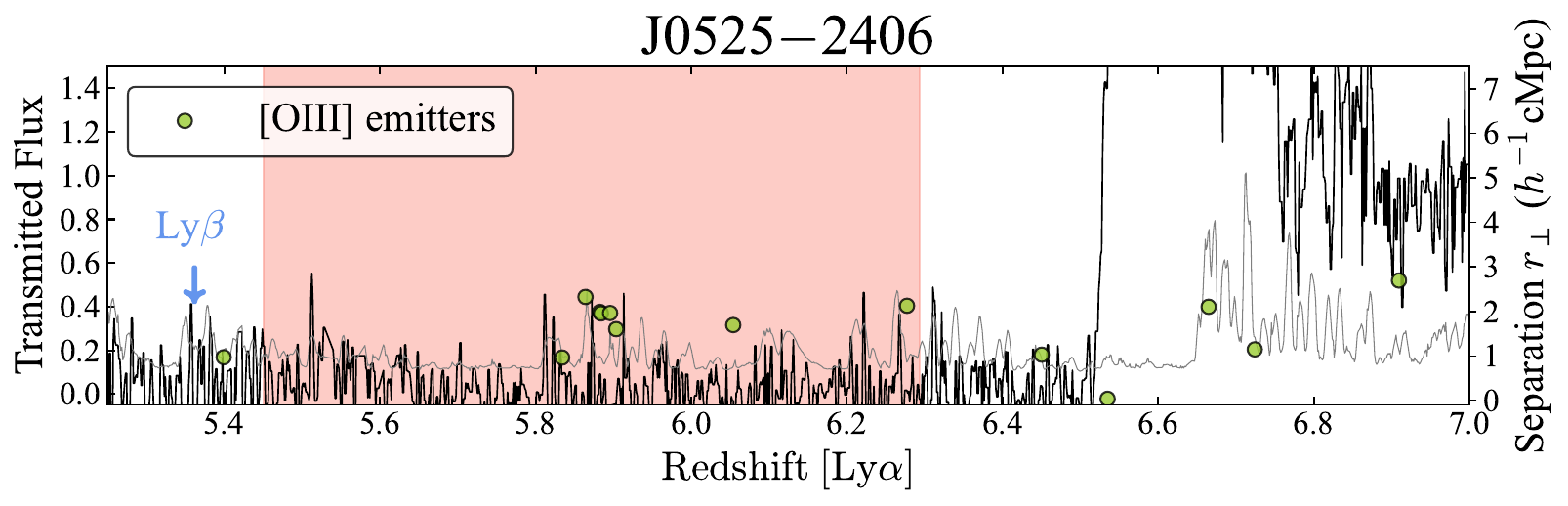}
    \includegraphics[width=0.95\textwidth]{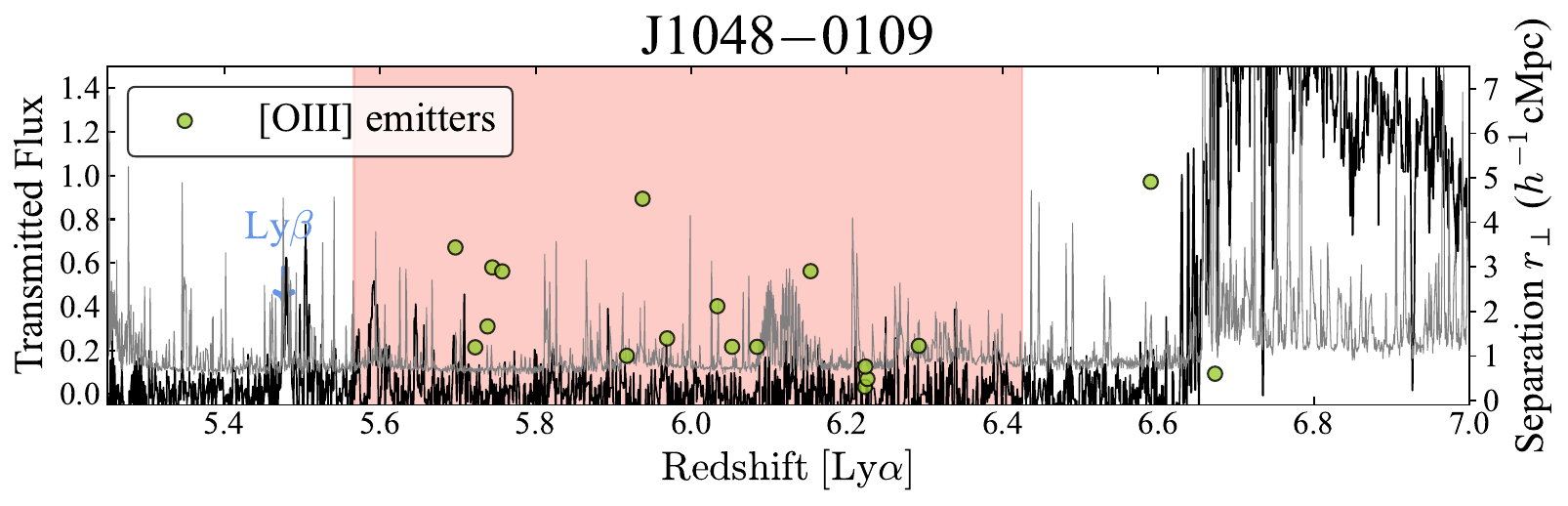}
    \caption{(continued.) The transmitted spectrum of ASPIRE quasars J0244$+$5008, J0305$-$3150, J0525$-$2406, and J1048$-$0109, together with \oiii\ emitters detected in the quasar fields.}
    \label{fig:los_plot_2}
\end{figure*}

\begin{figure*}[!ht]
    \centering
    \includegraphics[width=0.95\textwidth]{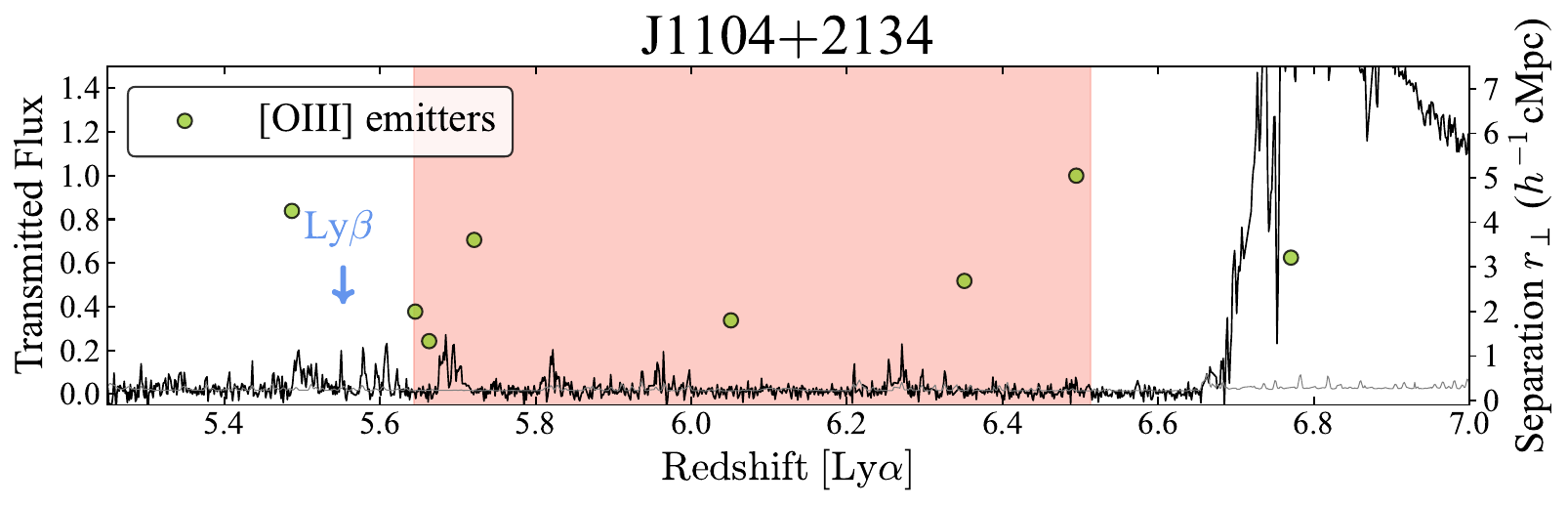}
    \includegraphics[width=0.95\textwidth]{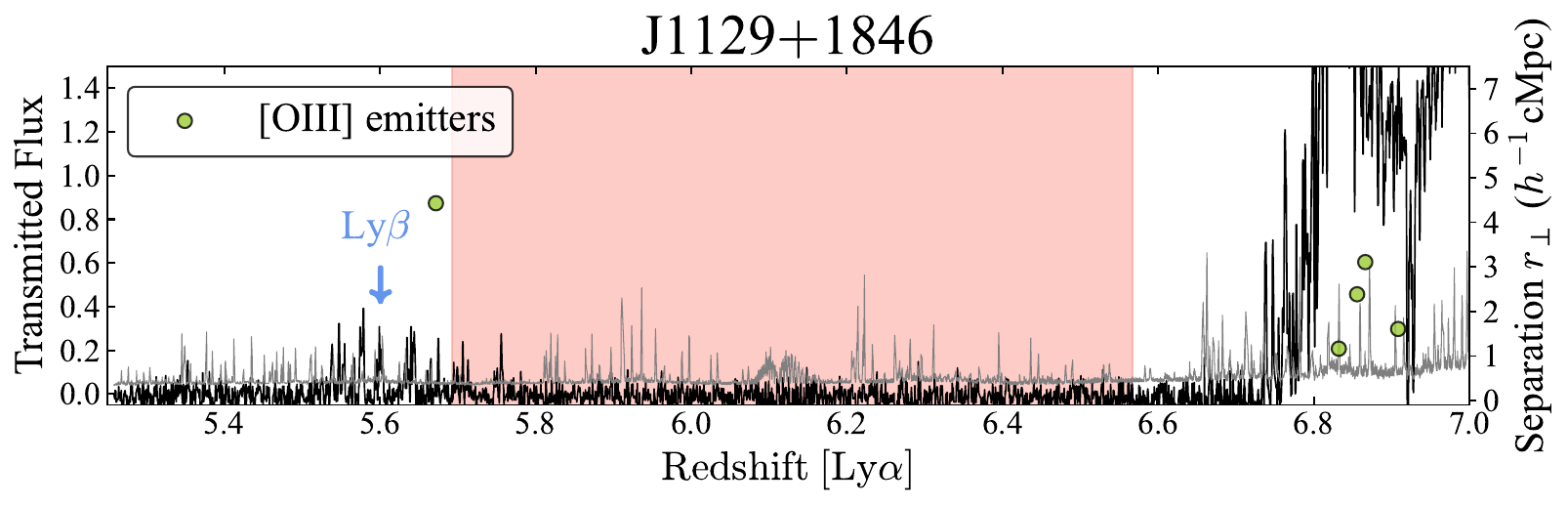}
     \includegraphics[width=0.95\textwidth]{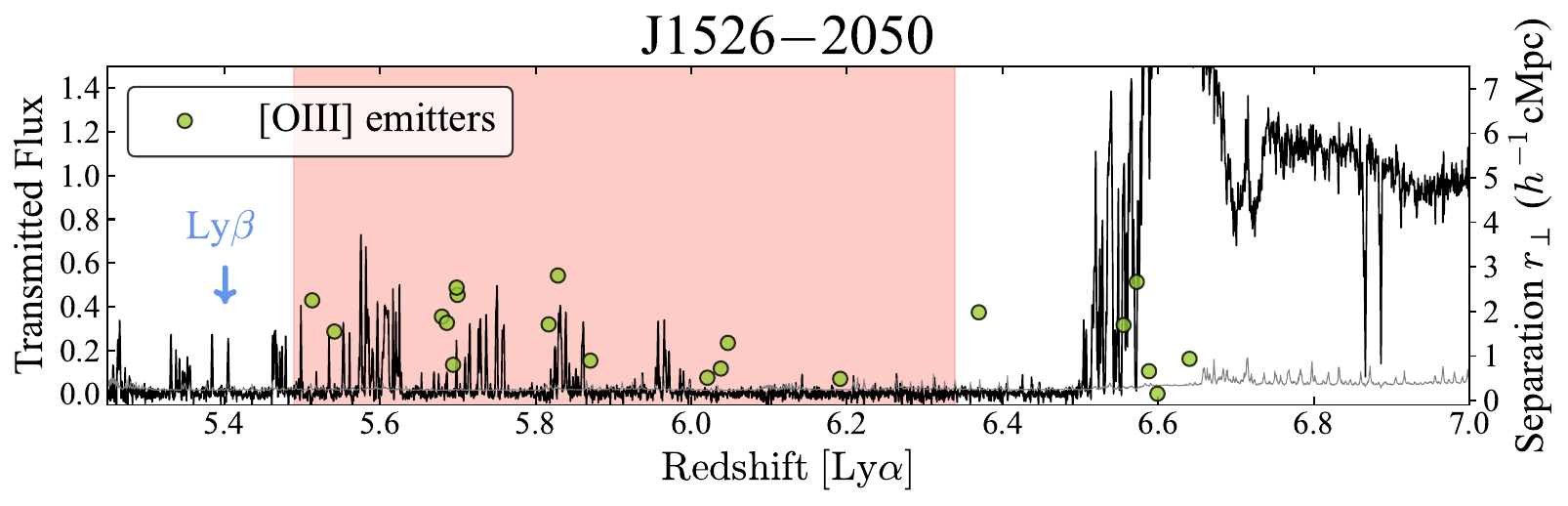}
    \includegraphics[width=0.95\textwidth]{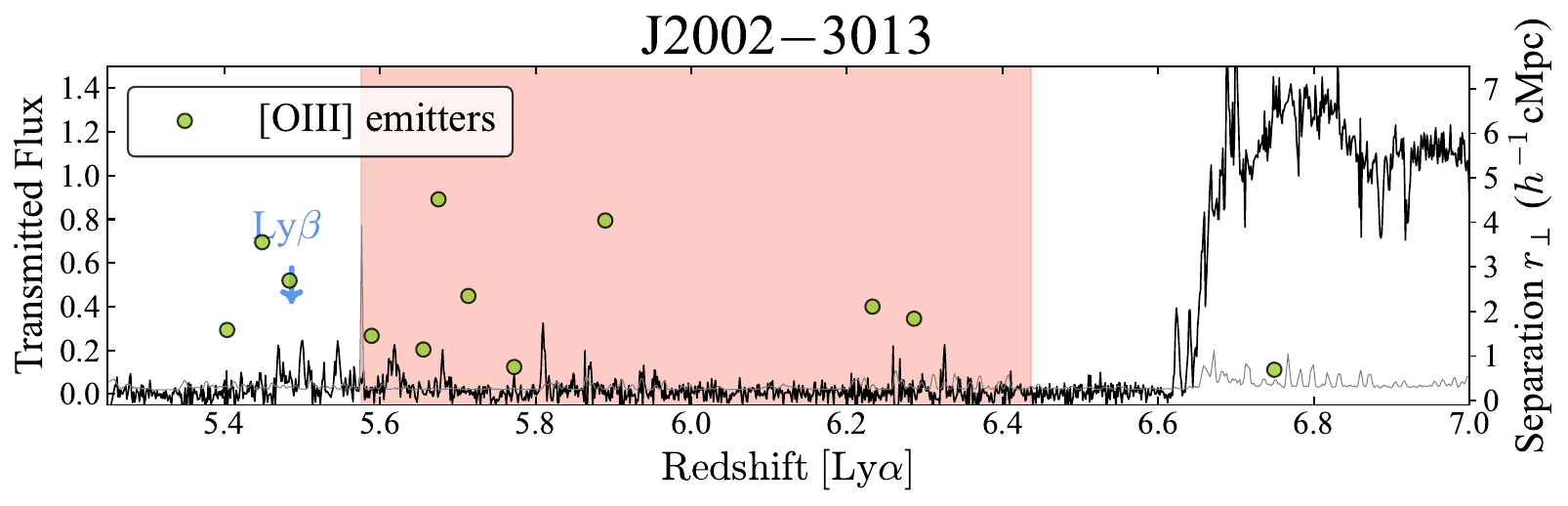}
    \caption{(continued.) The line of sight plot of ASPIRE quasars J1104$+$2134, J1129$+$1846, J1526$-$2050, and J2002$-$3013, together with \oiii\ emitters detected in the quasar fields.}
    \label{fig:los_plot_3}
\end{figure*}

\begin{figure*}[!ht]
    \centering
   
    \includegraphics[width=0.95\textwidth]{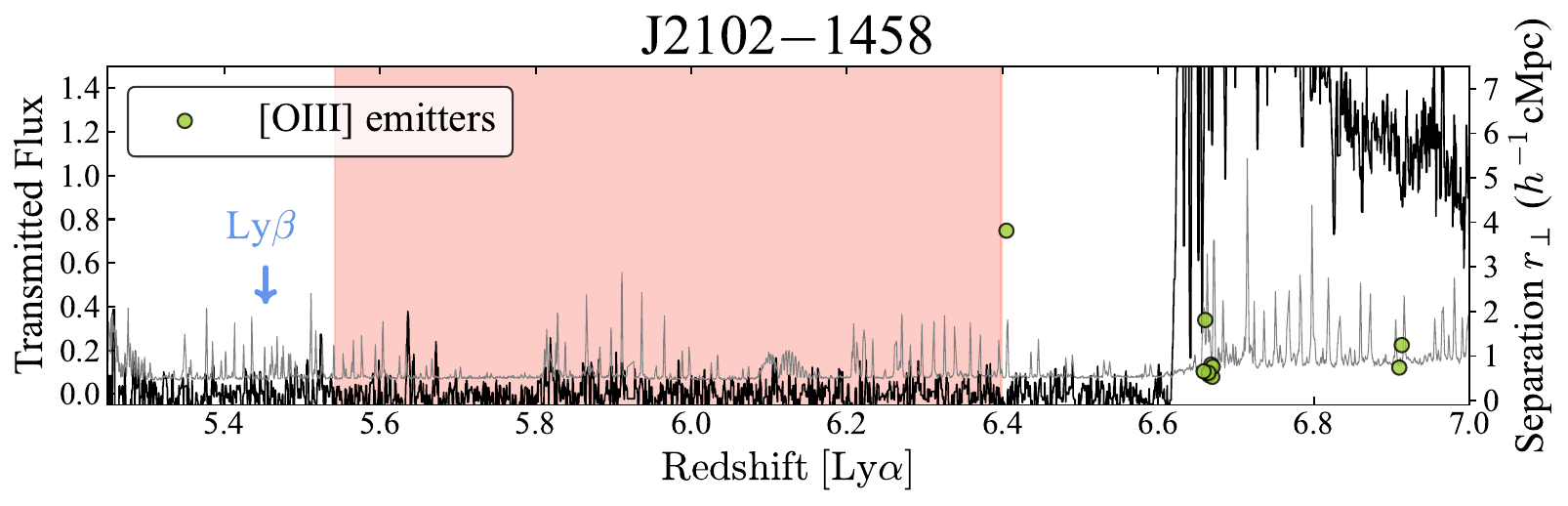}
    \includegraphics[width=0.95\textwidth]{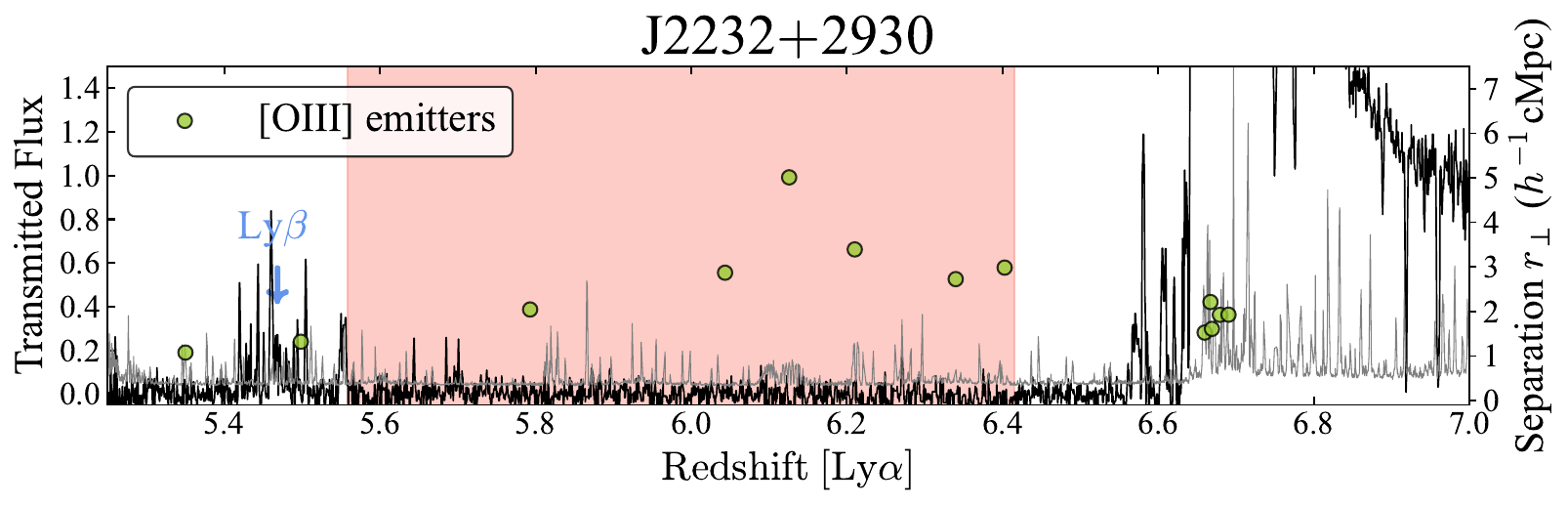}
    \caption{(continued.) The line of sight plot of ASPIRE quasars J2102$-$1458 and J2232$+$2930, together with \oiii\ emitters detected in the quasar fields.}
    \label{fig:los_plot_4}
\end{figure*}

\section{Consistency Check between the Random $\tau_{\rm eff,[OIII]}$ Distribution and the IGM Effective Optical Depth Distribution}\label{appendix:random_check}

In this section, we investigate whether the random distribution of $\tau_{\rm eff}$ around \oiii\ emitters ($\tau_{\rm eff,[OIII]}$) can recover the $\tau_{\rm eff}$ distribution $\tau_{\rm eff}$ of IGM patches at the corresponding redshifts. For the Ly$\alpha$ forest spectrum of each quasar, we use the spatial locations (redshift and transverse distance) of \oiii\ emitters detected in the other 13 quasar fields to compute the random $\tau_{\rm eff,[OIII]}$, following the method described in Section \ref{sec:method}. For each influence radius, we use twice the influence radius as the bin size to calculate the $\tau_{\rm eff}$ of IGM patches, starting from the rest frame 1040~\AA\ up to the rest frame 1176~${\rm \AA}$ in the quasar Ly$\alpha$ forest. 

\begin{figure*}
    \centering
    \includegraphics[width=1.0\textwidth]{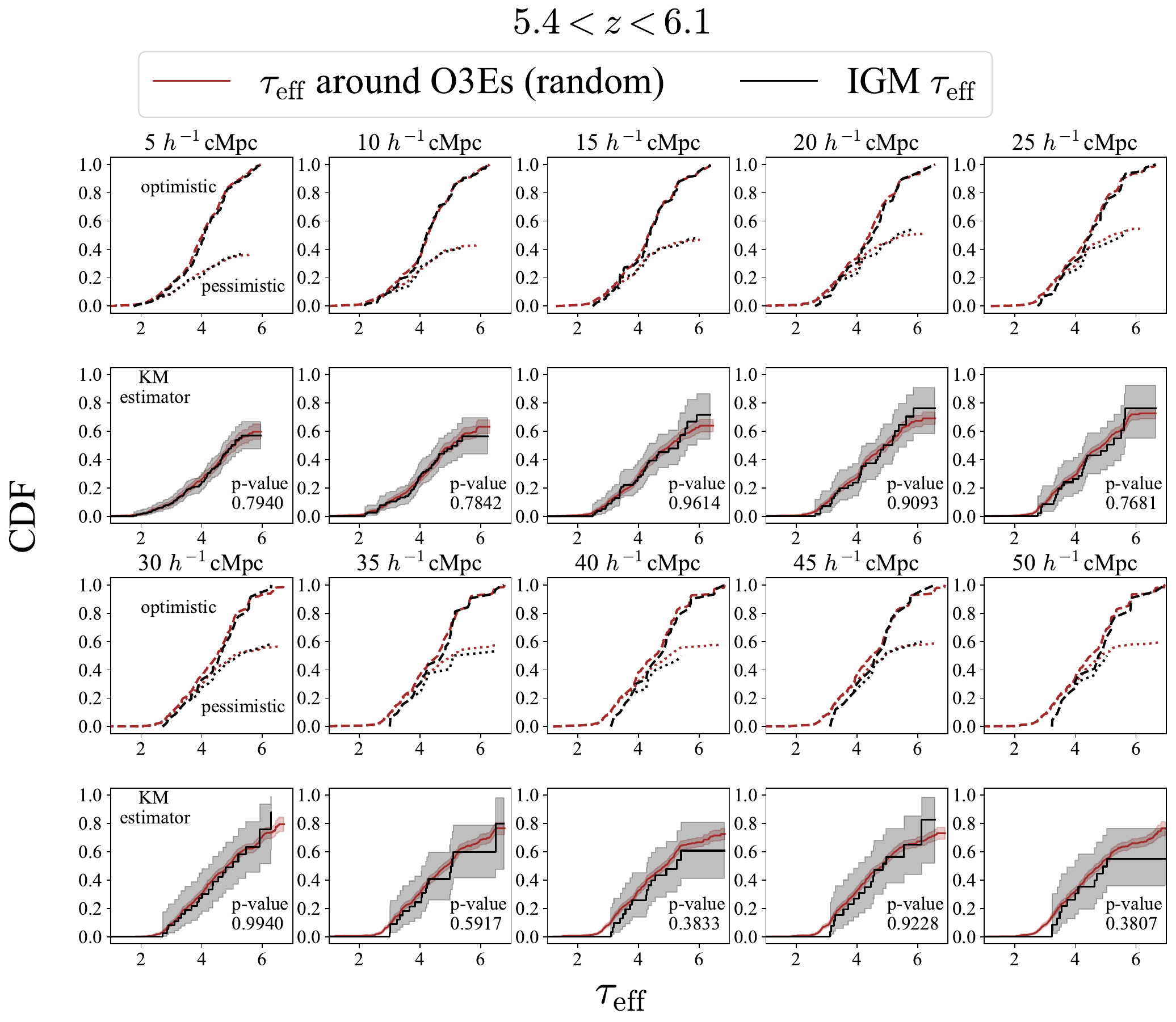}
    \caption{The first and the third rows show the cumulative distribution functions of the random $\tau_{\rm eff}$ around \oiii\ emitters (brown), and $\tau_{\rm eff}$ of IGM patches (black) at $5.4<z<6.1$. The influence radius is from 5~$h^{-1}\;\!{\rm cMpc}$ to 50~$h^{-1}\;\!{\rm cMpc}$ for $\tau_{\rm eff,[OIII]}$ measurements and $\tau_{\rm eff}$ are measured using the bin size of twice the corresponding influence radius. The ``optimistic" and the ``pessimistic" CDFs are plotted in dashed and dotted lines, respectively. The second and the fourth row show the fitted CDFs from Kaplan-Meier (KM) estimator and $1\sigma$ confidence interval of the CDF is shown in the shaded regions of the corresponding color. The null-hypothesis p-value from the log-rank test is shown in the bottom right corner of each sub-panel. With all influence radii, the random $\tau_{\rm eff,[OIII]}$ distribution is consistent with $\tau_{\rm eff}$ distribution of IGM patches.}
    \label{fig:random_cdf_different scales}
\end{figure*}

\begin{figure*}
    \centering
    \includegraphics[width=1.0\textwidth]{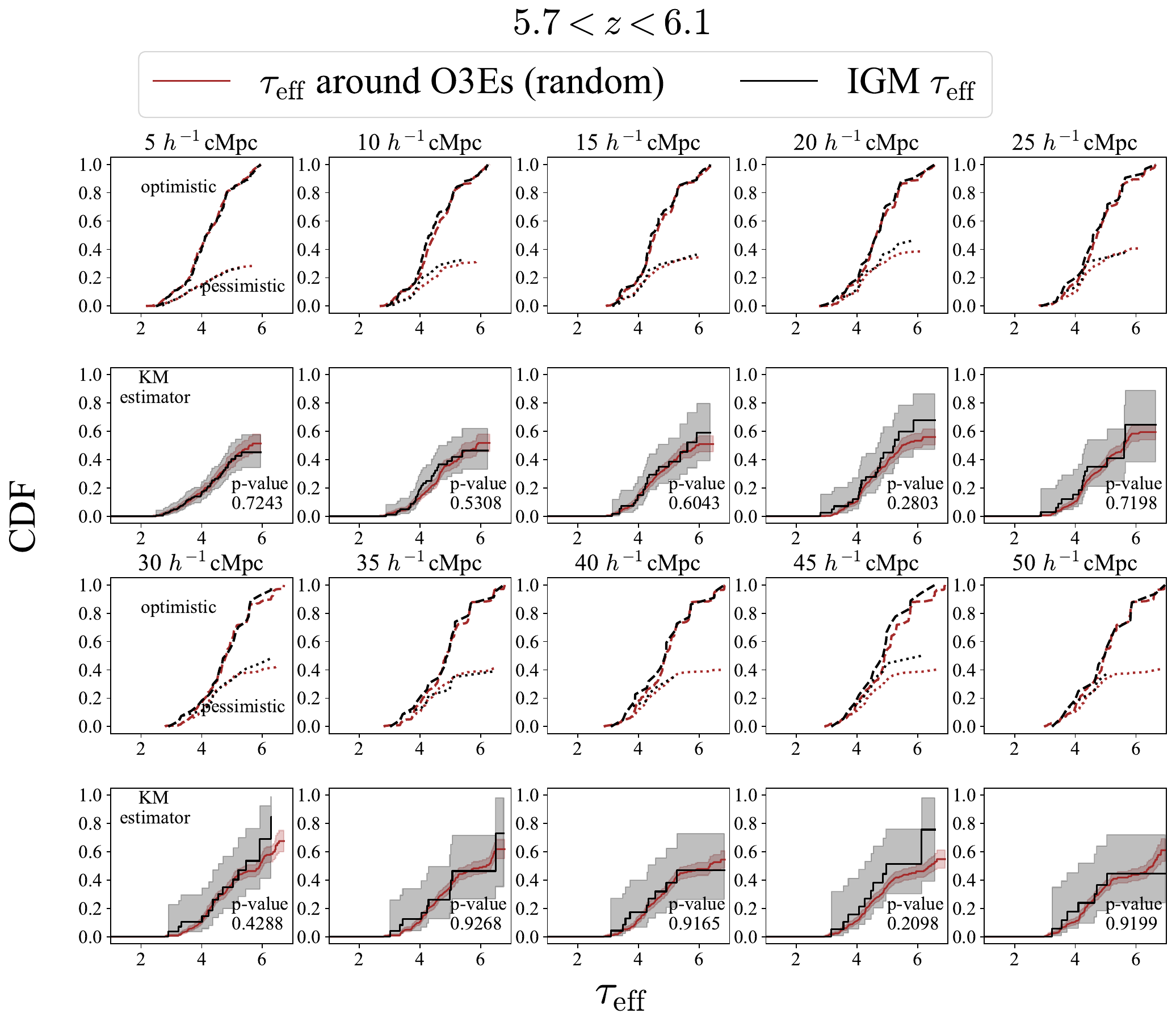}
    \caption{The cumulative distribution function of the random $\tau_{\rm eff,[OIII]}$ (brown) and $\tau_{\rm eff}$ of IGM patches (black) at $5.7<z<6.1$.}
    \label{fig:random_cdf_different scales_highz}
\end{figure*}

\begin{figure*}
    \centering
    \includegraphics[width=1.0\textwidth]{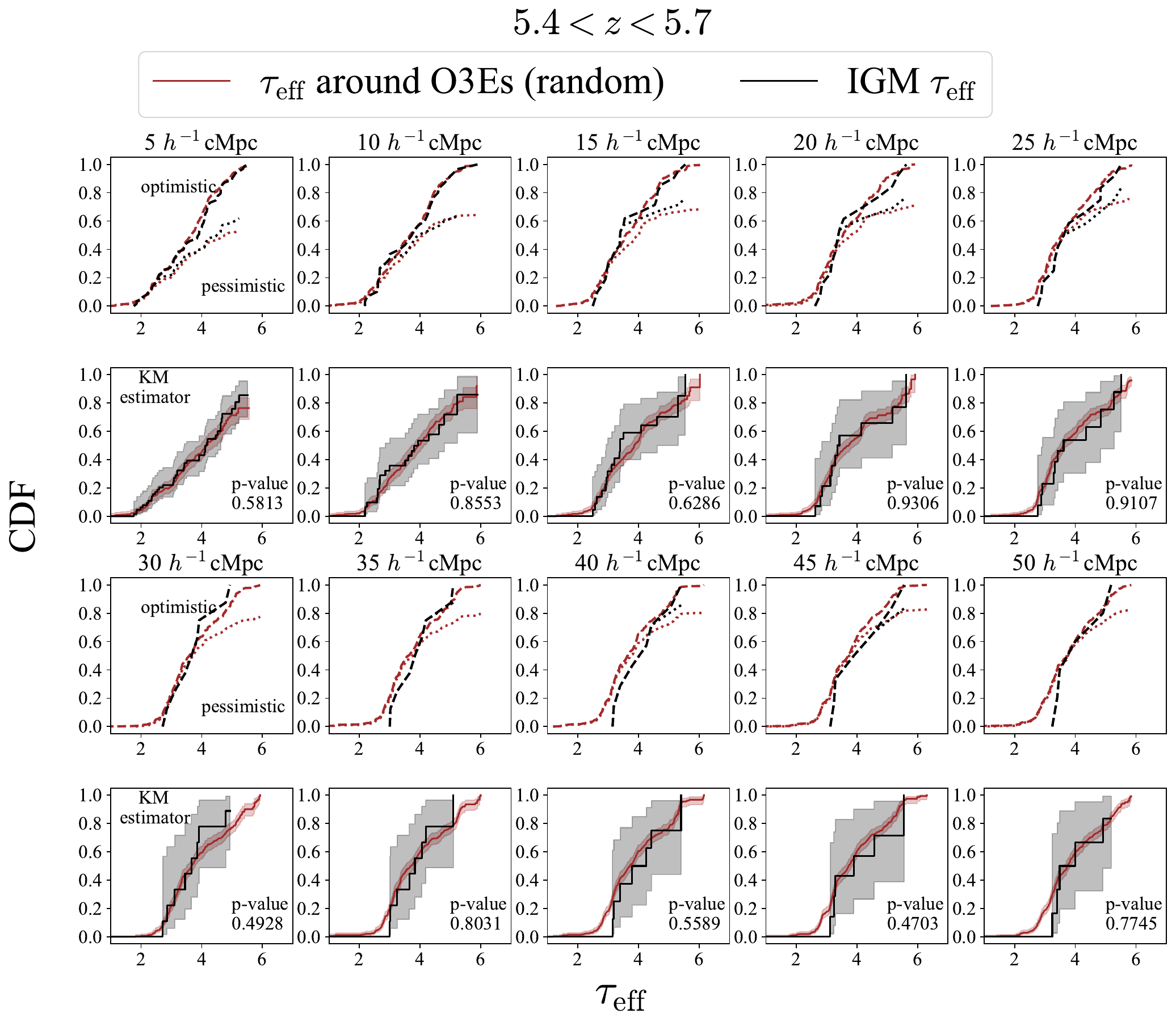}
    \caption{The cumulative distribution function of the random $\tau_{\rm eff,[OIII]}$ (brown) and $\tau_{\rm eff}$ of IGM patches (black) at $5.4<z<5.7$.}
    \label{fig:random_cdf_different scales_lowz}
\end{figure*}

We plot the ``optimistic" CDFs of random $\tau_{\rm eff,[OIII]}$ distribution and the $\tau_{\rm eff}$ distribution of IGM patches in brown and black dashed lines. The corresponding ``pessimistic" CDFs are denoted by the dotted lines. 
We then use the KM estimator to fit the survival function of each distribution. We show the fitted CDFs and $1\sigma$ confidence intervals of random $\tau_{\rm eff,[OIII]}$ distribution and $\tau_{\rm eff}$ distribution of IGM patches by brown and black solid lines and shaded regions in Figure \ref{fig:random_cdf_different scales} ($5.4<z<6.1$), Figure \ref{fig:random_cdf_different scales_highz} ($5.7<z<6.1$), and Figure \ref{fig:random_cdf_different scales_lowz} ($5.4<z<5.7$), respectively. We perform the log-rank test between these two distributions and p-values for null-hypothesis are greater than 0.05 for any influence radius in all three redshift bins, indicating the random $\tau_{\rm eff,[OIII]}$ distribution is consistent with the $\tau_{\rm eff}$ distribution of IGM patches. As such, the random $\tau_{\rm eff,[OIII]}$ distribution can well represent the $\tau_{\rm eff}$ distribution of IGM patches at the corresponding redshifts. 

At $\tau_{\rm eff}\lesssim3$ in both $5.4<z<6.1$ and $5.4<z<5.7$ bins, the random $\tau_{\rm eff,[OIII]}$ distribution tends to have a more extended low $\tau_{\rm eff}$ tail than the $\tau_{\rm eff}$ distribution of IGM patches. This is due to the fact that our method for measuring the $\tau_{\rm eff,[OIII]}$ can include \oiii\ emitters with redshift less than the low redshift cut of the Ly$\alpha$ forest, as long as the \oiii\ emitters can enclose part of the Ly$\alpha$ forest spectrum within the influence radius. Equivalently, $\tau_{\rm eff,[OIII]}$ include the spectrum close to the boundary of the Ly$\alpha$ forest, while the $\tau_{\rm eff}$ distribution of IGM patches is calculated within a large bin size with a midpoint redshift slightly higher than the low redshift boundary of the Ly$\alpha$ forest. As the $\tau_{\rm eff}$ increases with redshift, the $\tau_{\rm eff}<3$ cannot be well sampled by the $\tau_{\rm eff}$ distribution of IGM patches. 

Furthermore, because the number of quasar sightlines is only 14, the number of IGM patches is very limited when adopting a large bin size to calculate $\tau_{\rm eff}$. This effect becomes more obvious at $5.4<z<5.7$ than at $5.7<z<6.1$ since not every quasar sightline fully covers the Ly$\alpha$ forest at $5.4<z<5.7$. Therefore, we use the random $\tau_{\rm eff,[OIII]}$ distribution as the control sample for the IGM transmission to compare with the $\tau_{\rm eff,[OIII]}$ distribution. 

%% For this sample we use BibTeX plus aasjournals.bst to generate the
%% the bibliography. The sample631.bib file was populated from ADS. To
%% get the citations to show in the compiled file do the following:
%%
%% pdflatex sample631.tex
%% bibtext sample631
%% pdflatex sample631.tex
%% pdflatex sample631.tex

\bibliography{sample631}{}
\bibliographystyle{aasjournal}

%% This command is needed to show the entire author+affiliation list when
%% the collaboration and author truncation commands are used.  It has to
%% go at the end of the manuscript.
%\allauthors

%% Include this line if you are using the \added, \replaced, \deleted
%% commands to see a summary list of all changes at the end of the article.
%\listofchanges

\end{document}